\documentclass[conference]{IEEEtran}
\IEEEoverridecommandlockouts
\usepackage{cite}
\usepackage{amsmath,amssymb,amsfonts}
\usepackage{algorithmic}
\usepackage{graphicx}
\usepackage{textcomp}
\usepackage[dvipsnames]{xcolor}

\usepackage{multirow}
\usepackage[shortlabels]{enumitem}
\usepackage{graphicx}
\usepackage{subfigure}
\usepackage{amsmath}
\usepackage{float}
\usepackage{ulem} 

\newcommand{\carl}[1]{\textcolor{black}{#1}}
\newcommand{\mbx}{\boldsymbol{x}}
\newcommand{\mbu}{\boldsymbol{u}}
\newcommand{\mbv}{\boldsymbol{v}}
\newcommand{\mbU}{\boldsymbol{U}}
\newcommand{\mbV}{\boldsymbol{V}}

\newcommand{\eg}{\textit{e.g.}}
\newcommand{\ie}{\textit{i.e.}}

\newcommand{\wrt}{\textit{w.r.t.~}}

\begin{document}

\title{Multi-Facet Recommender Networks with \mbox{Spherical Optimization}}
\normalem

\author{
Yanchao Tan$^{\dagger}$, Carl Yang$^{\ddagger}$, Xiangyu Wei$^{\dagger}$, Yun Ma$^{\dagger}$, Xiaolin Zheng$^{\dagger*}$\thanks{*Corresponding author}
\\
\normalsize $^{\dagger}$\emph{College of Computer Science, Zhejiang University, Hangzhou, China} \\
\normalsize $^{\ddagger}$\emph{Department of Computer Science, Emory University, Atlanta, United States} \\
\emph{$^{\dagger}$\{yctan, weixy, 3170105433, xlzheng\}@zju.edu.cn ~~~~~ $^{\ddagger}$j.carlyang@emory.edu} \\
}

\setlength{\floatsep}{4pt plus 4pt minus 1pt}
\setlength{\textfloatsep}{4pt plus 2pt minus 2pt}
\setlength{\intextsep}{4pt plus 2pt minus 2pt}
\setlength{\dbltextfloatsep}{3pt plus 2pt minus 1pt}
\setlength{\dblfloatsep}{3pt plus 2pt minus 1pt}
\setlength{\abovecaptionskip}{3pt}
\setlength{\belowcaptionskip}{2pt}
\setlength{\abovedisplayskip}{2pt plus 1pt minus 1pt}
\setlength{\belowdisplayskip}{2pt plus 1pt minus 1pt}

\maketitle

\begin{abstract}
Implicit feedback is widely explored by modern recommender systems. Since the feedback is often sparse and imbalanced, it poses great challenges to the learning of complex interactions among users and items. Metric learning has been proposed to capture user-item interactions from implicit feedback, but existing methods only represent users and items in a single metric space, ignoring the fact that users can have multiple preferences and items can have multiple properties, which leads to potential conflicts limiting their performance in recommendation. To capture the multiple facets of user preferences and item properties while resolving their potential conflicts, we propose the novel framework of Multi-fAcet Recommender networks with Spherical optimization (MARS). By designing a cross-facet similarity measurement, we project users and items into multiple metric spaces for fine-grained representation learning, and compare them only in the proper spaces. Furthermore, we devise a spherical optimization strategy to enhance the effectiveness and robustness of the multi-facet recommendation framework.
Extensive experiments on six real-world benchmark datasets show drastic performance gains brought by MARS, which constantly achieves up to 40\% improvements over the state-of-the-art baselines regarding both HR and nDCG metrics.\footnote{\carl{https://github.com/Melinda315/MARS}}
\end{abstract}

\section{Introduction}
\begin{figure*}
    \centering
    \subfigure[]
    {
        \includegraphics[width=0.43\linewidth]{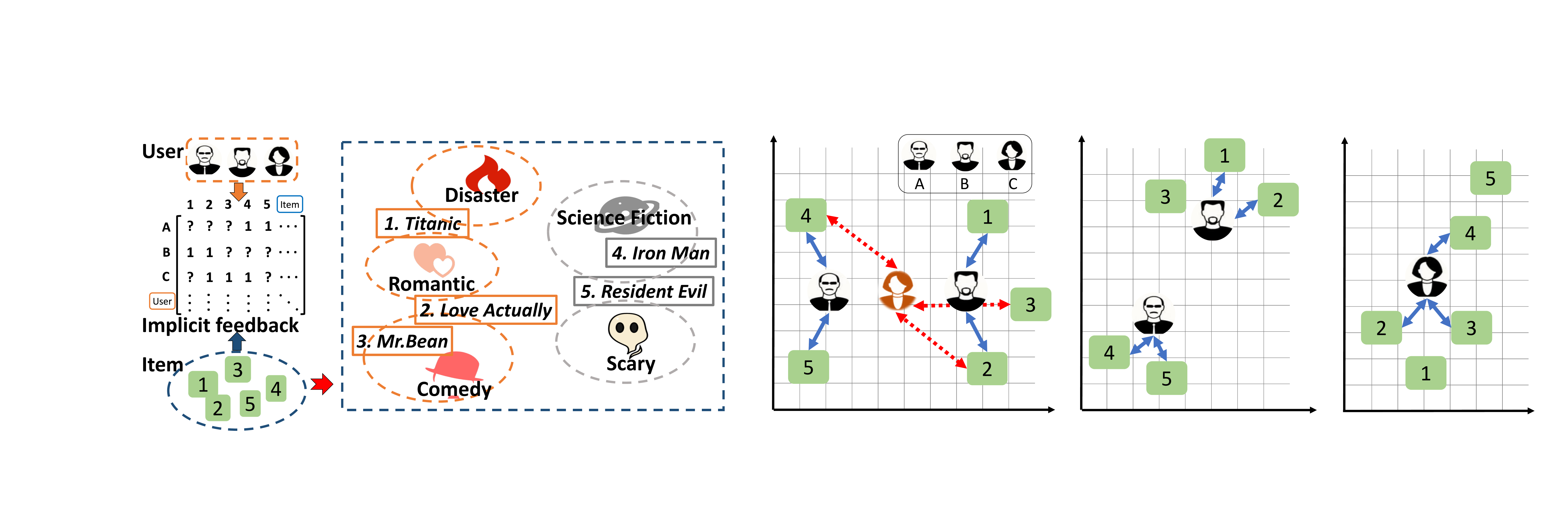}
        \label{fig:a}
    }
    \subfigure[]
    {
        \includegraphics[scale=0.16]{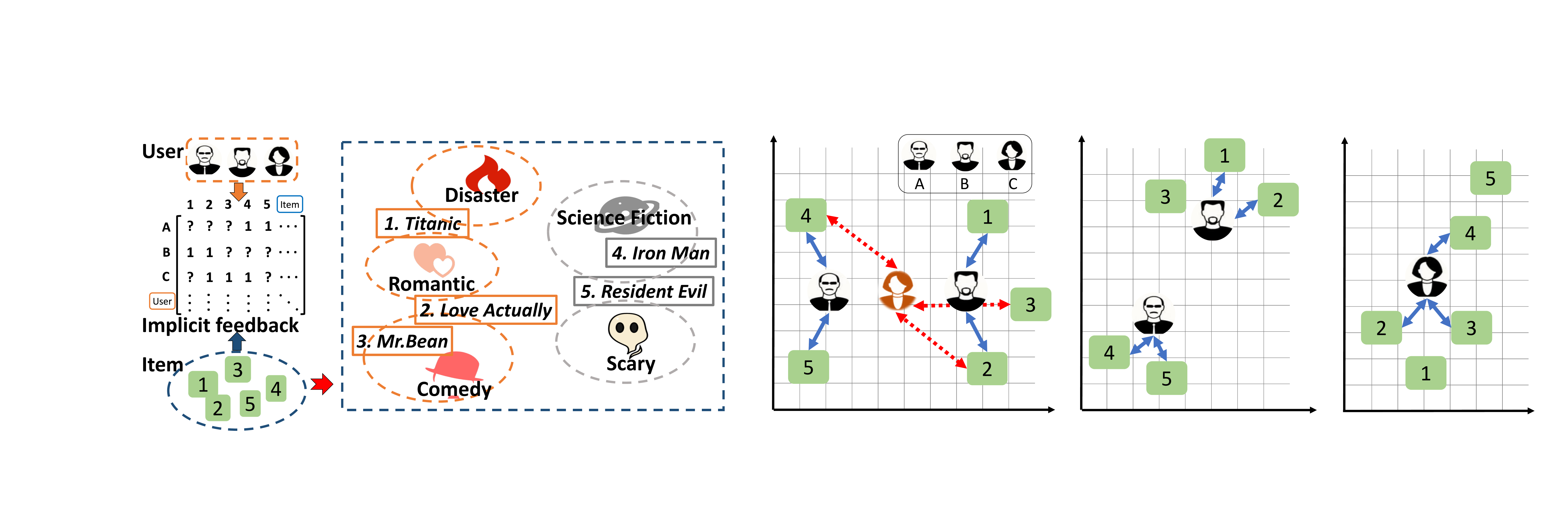}
        \label{fig:b}
    }
    \subfigure[]
    {
        \includegraphics[scale=0.16]{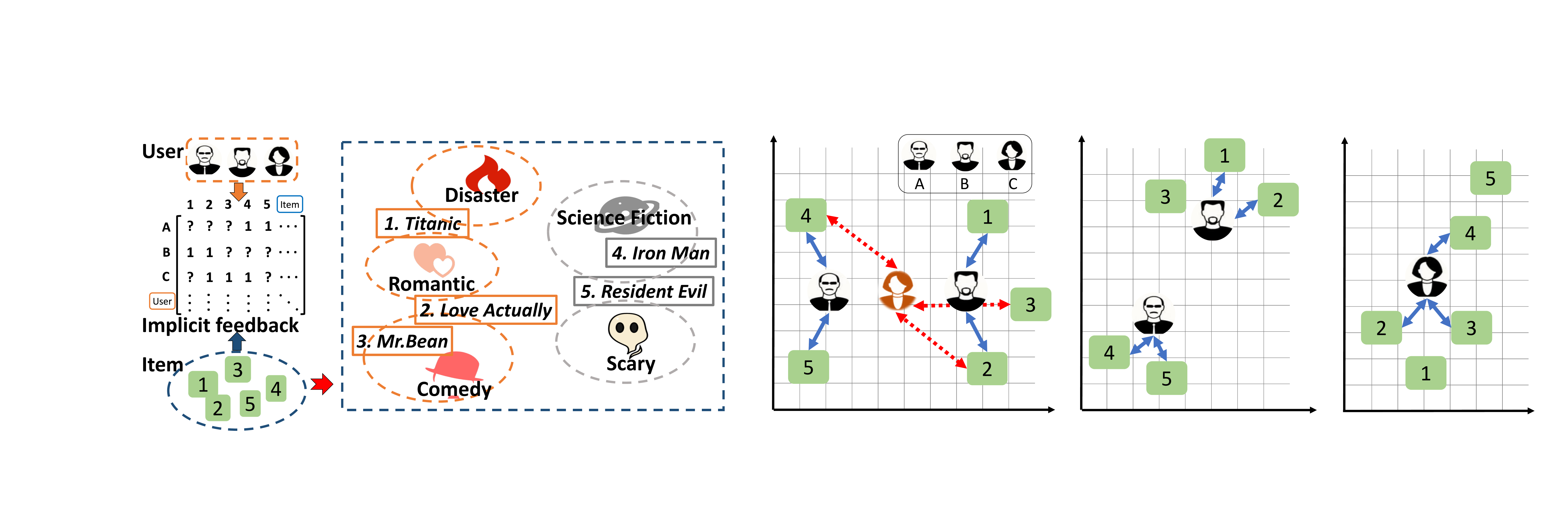}
        \label{fig:c}
    }
    \caption{A toy example illustrating the limitation of single-facet recommendation. (a) shows the implicit feedback data and their underlying multi-facet distribution to be modeled by a recommender system. (b) shows users and items in a single metric space. Assuming users (A, B) and items (1-5) are arranged in the black and green locations, where the blue arrows faithfully preserve the interactions between users (A, B) and items (1-5), it is then impossible to put user C anywhere in that space without violating her interactions with items (2-4) as shown by the red arrows.
    (c) shows the same users and items in multiple metric spaces, which implicitly captures user-item interactions regarding different facets and naturally overcomes the limitation of (b).}
    \label{fig:multi}
\end{figure*}

With the rapid growth of various activities on the Web, recommender systems become fundamental in helping users alleviate the problem of information overload.
Compared with explicit feedbacks (\eg, 1-5 star ratings), implicit feedbacks (\eg, purchase records and browsing history, as shown in Figure \ref{fig:a}) are much more abundant and accessible in real-world online applications \cite{joachims2017accurately}. 
Due to the extreme sparsity and imbalance of implicit feedbacks, several methods based on metric learning \cite{hsieh2017collaborative, park2018collaborative, tay2018latent, zhang2018metric} have been proposed recently, which have been shown advantageous over the classic matrix factorization based methods \cite{hu2008collaborative, rendle2009bpr, zhao2015improving, weston2010large, he2018adversarial}.

However, in many real-world applications, users can have multiple preferences and items can have multiple properties (uniformally termed as multi-facet). As an example, in Figure \ref{fig:a}, the movie \textit{Love Actually} belongs to both \textit{romantic} and \textit{comedy} categories, so user B may be attracted by its romantic plots, whereas user C by its humorous actors. 
By using a single metric space to project all users and items, existing metric learning based recommender systems ignore the possible multi-facet user preferences and item properties. 

\carl{As shown in Figure \ref{fig:b}, the preference of user C would require items 2 and 4 to be close, while those of users A and B would require items 2 and 4 to stay away from each other, leading to an unresolvable conflict in the single embedding space. Particularly, the two items cannot simultaneously be close and far away regardless of the embedding dimension. Our solution is to allow users and items to reside in multiple embedding spaces and compute their similarity only in the proper spaces. In this way, the two items can be close in one space (corresponding to user C’s preference), while far away in another (corresponding to the preference of users A and B), which effectively resolves the potential conflicts caused by multi-facet user preferences and item properties (as shown in Figure \ref{fig:c}.}

In this work, we enable such a framework of Multi-fAcet Recommender networks (MAR), whose major goal is to simultaneously learn multiple expressive metric (embedding) spaces of users and items directly from the implicit feedback data (Section \ref{sec:mar:a}).
The task is challenging in several perspectives.

Firstly, although multi-space representation learning has been explored very recently \cite{epasto2019single, liu2019single, wang2019mcne, jiang2019role}, existing works have only studied it with traditional network data instead of recommendation data, which is close to an extremely sparse bipartite network. Moreover, they learn different embedding spaces under the help of additional graph clustering algorithms \cite{epasto2019single, liu2019single}, auxiliary category information \cite{wang2019mcne}, or pre-defined textual patterns \cite{jiang2019role}. 
Instead of relying on any of those external help, which is often unavailable in recommender systems, we propose a \textit{cross-facet similarity measurement}, which naturally connects multi-space representation learning with metric learning based recommendation, and allows the simultaneous learning of multiple facet-specific embedding spaces for users and items in a both memory and computation efficient way (Section \ref{sec:mar:b}).

Secondly, the standard optimization objectives of recommender systems are not designed towards the learning of multiple metric spaces, where the key idea is to properly combine and fully utilize the representation power of multiple spaces. To this end, we design a series of \textit{multi-facet optimization objectives}, most of which are inspired by existing works in other lines of research such as CV \cite{sun2020circle} and social science \cite{komiak2006effects}, and novelly integrate them as a whole (Section \ref{sec:mar:c}).

Finally, we further find flaws in the standard constraints of metric learning on the norms of user and item embeddings. We deem the norm constraints not strong enough and prone to trivial optimization with difficult users and items, which is especially concerning in the multi-space setting as more representation power can be wasted. In light of this, we devise a spherical optimization strategy based on recent research on machine learning for NLP \cite{meng2019spherical} and CV \cite{ranjan2018crystal}, which strictly constrains the norms of user and item embeddings in all facet-specific metric spaces. The whole MAR with Spherical optimization framework is named MARS (Section \ref{sec:mars}).

We evaluate both MAR and MARS with experiments on \carl{six} real-world benchmark datasets for recommendation with implicit feedback. We compare them with a comprehensive set of eight recommendation methods focusing on the classic and state-of-the-art metric learning based methods. Extensive experimental results show that MAR alone is able to significantly improve the recommendation over all baselines (\eg, with \carl{up to 27.53\%} relative improvements on HR@10 over the best baselines across all datasets), whereas MARS can further drastically improve over MAR especially in the harder cases (\eg, with \carl{up to 47.07\%} relative improvements on HR@10 over the best baselines). More comprehensive results and discussions as well as ablation study, hyperparameter study and case study are all presented in Section \ref{sec:exp}.

 \section{Related Work}
\subsection{Recommender systems}
Over the past decade, matrix factorization (MF) has become the \textit{de facto} method for recommendation, which uses inner products to model the similarity of user-item pairs \cite{paatero1994positive, rendle2009bpr, hu2008collaborative, yang2017bridging, liu2020certifiable}.
However, MF assumes linear relationships between users and items, which limits the model capacity, since the interactions between users and items in real-world applications are often much more complex. 
In particular, the inner product used by MF cannot satisfy the triangle inequality, which limits the capabilities of the models to capture fine-grained user-user and item-item similarities \cite{hsieh2017collaborative}.

Recently, metric learning for recommendations has attracted significant research attention \cite{weinberger2009distance,yang2006distance,zadeh2016geometric,song2017parameter,hsieh2017collaborative}. Existing methods in this line seek appropriate distance functions for input points instead of inner products, which can address the limitations of MF. 
Based on the Euclidean distance, \cite{hsieh2017collaborative} first proposed a method called collaborative metric learning (CML), which learns a metric space to encode not only users’ preferences but also the user-user and item-item similarity. 
To avoid pushing possible recommendation candidates too far away, \cite{zhang2018metric} proposed a metric factorization method only with the pulling operation in contrast to CML, which only has a pushing term. 
Since CML has a one-to-many mapping problem which limits the representation of users and items, \cite{park2018collaborative} turned this problem to multiple one-to-one mappings and \cite{DBLP:conf/wsdm/TranT0CL20} turned this problem to one-to-one mappings between Euclidean and hyperbolic spaces. Then, inspired by the success of relational metric learning in knowledge graphs \cite{wang2014knowledge}, it constructed user-item translation vectors by employing the neighborhood information of users and items. In this case, the model can not only push the negative items away from the user but also pull a user closer to all of the interacted items. 
Considering that CML has a geometrically restrictive scoring function and it has been proven to be an ill-posed algebraic system, \cite{tay2018latent} learned latent user-item interaction relations based on memory network and attention mechanism, which helps to alleviate the potential geometric problem.

Although the above single-space metric learning methods achieve promising performance, the representation capacity of a single space is limited and they cannot explicitly decompose the multiple facets and similarities of users and items. 
Recent research on network representation learning has shown the necessity and effectiveness of multi-space embedding \cite{chen2018pme, liu2019single, epasto2019single, wang2019mcne, jiang2019role}.
\carl{The idea is extensible to recommender systems, where the multiple spaces can be naturally used to model the multi-facet user preferences and item properties.}
\carl{Beyond matrix factorization and metric learning, recent recommendation methods like \cite{klyuchnikov2019figuring} leveraged cokriging for active item retrieval to capture user's multiple interests under minimal user input, whereas \cite{MultiSage} adopted external context to explicitly model the multi-facet interactions among items to handle flexible user preferences. Their focuses on minimal user input and external context are different from ours, but they also show the potential of modeling multi-facet user preferences and item properties for effective recommendation.}

\subsection{Spherical optimization}
Most of the existing metric learning methods for recommendation only adopt the relaxed embedding norm constraints as CML \cite{hsieh2017collaborative}, which requires all user and item embeddings to lie in a unit sphere. However, there is no optimization for recommendation based on the strict spherical constraints, resulting in the discrepancy between the potentially more robust objective and its proper optimization.

Since the simultaneous modeling of multiple spaces will increase the learning burden, the optimization process has to be effective to fully exploit the increased representation capacity of the multi-space model. 
To properly optimize our model \wrt~the strict spherical constraints on embedding norms, we get inspired by recent advances in hyperspherical representation learning that have shown the effectiveness of Riemannian optimization in spherical spaces by focusing on the directional (cosine) similarity among embedding vectors \cite{meng2019spherical, ranjan2018crystal, shi2019probabilistic, mettes2019hyperspherical}.
For example, spherical generative modeling \cite{batmanghelich2016nonparametric, zhang2017triovecevent, zhuang2017identifying,huang2019mala} captures the distribution of words on the unit sphere, motivated by the effectiveness of directional metrics over word embeddings. 
Recently, spherical models also show great success in deep learning. Spherical normalization \cite{liu2017deep} on the input leads to easier optimization, faster convergence and better accuracy of neural networks, which helps regularize the vector against the input length and leads to better document clustering performance. Also, a spherical loss function can be used to replace the conventional softmax layer in language generation tasks, which results in faster and better generation quality \cite{kumar2018mises}. Motivated by the success of these models, we propose to directly optimize our recommender networks in the spherical space, with the properly designed calibrated Riemannian optimization which for the first time fills the gap between multi-space metric learning and spherical optimization. 

\section{The MAR Framework}
\label{sec:mar}
\begin{figure}[t!]
    \centering
    \includegraphics[width=\linewidth]{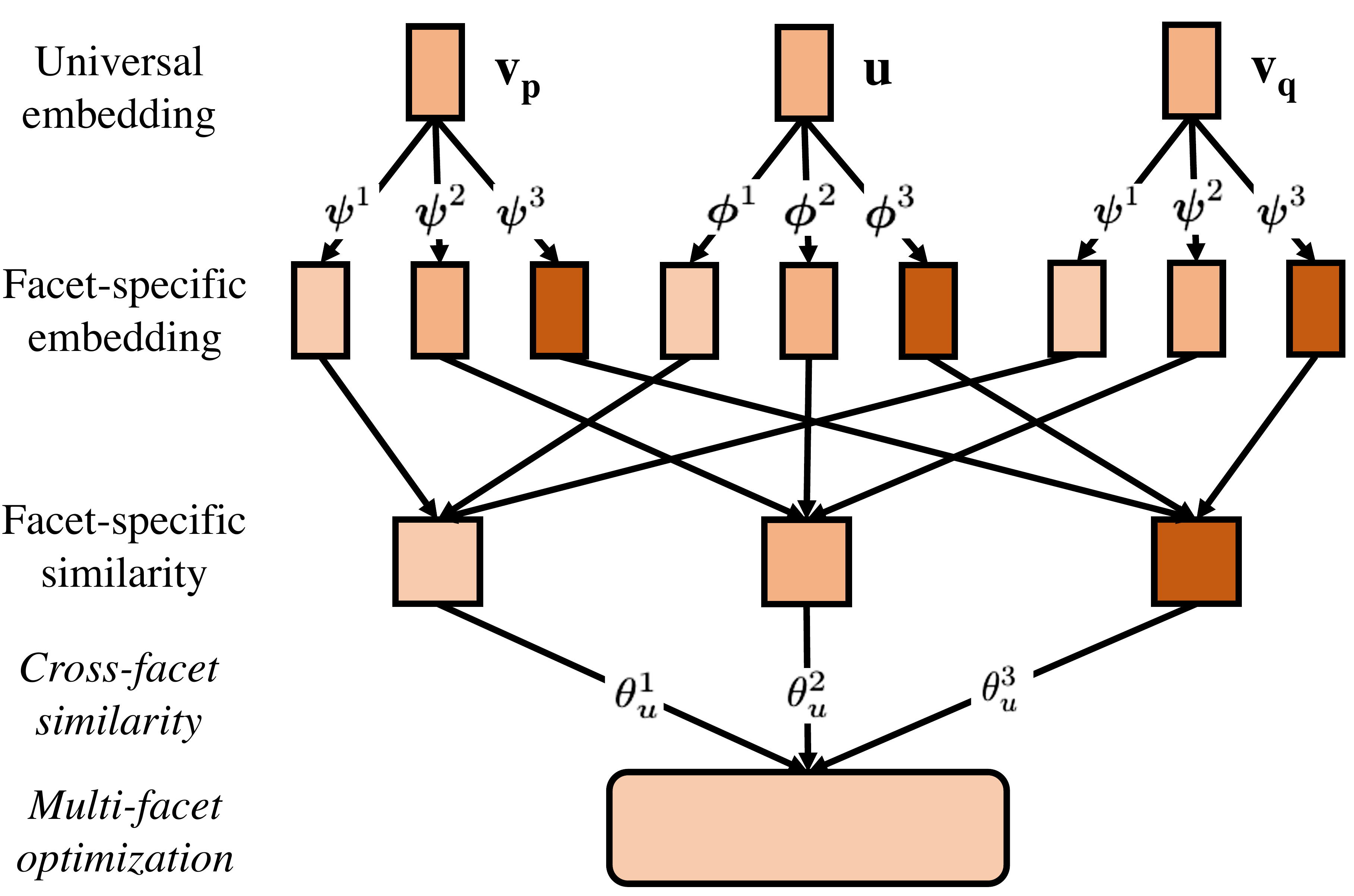}
    \caption{Overview of our proposed multi-facet recommender networks.}
    \label{fig:framework}
\end{figure}

In this section, we present our proposed framework of Multi-fAcet Recommender networks (MAR), as shown in Figure \ref{fig:framework}. Specifically, we firstly define a cross-facet similarity measurement, which projects users into multiple spaces to model their preferences from multiple perspectives. Then, we detail the optimization for MAR, which achieves effective personalized recommendation through the learning of multi-facet user and item embeddings.

\begin{figure*}
  \centering
    \subfigure[Single space recommendation]{
    \includegraphics[width = 0.35\linewidth]{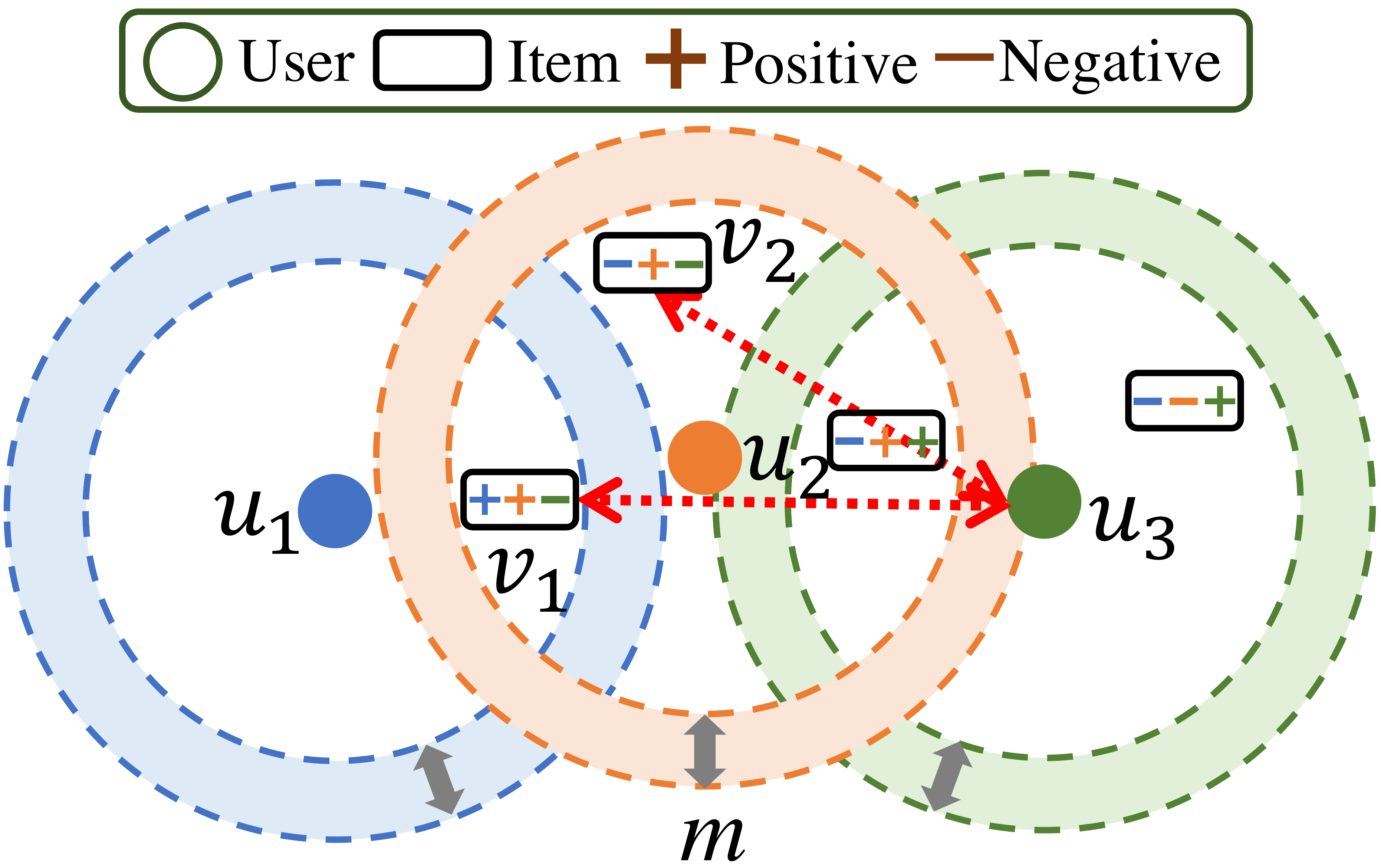}
    \label{fig:triplet}
    }
    \subfigure[Multi-space recommendation]{
    \includegraphics[width = 0.6\linewidth]{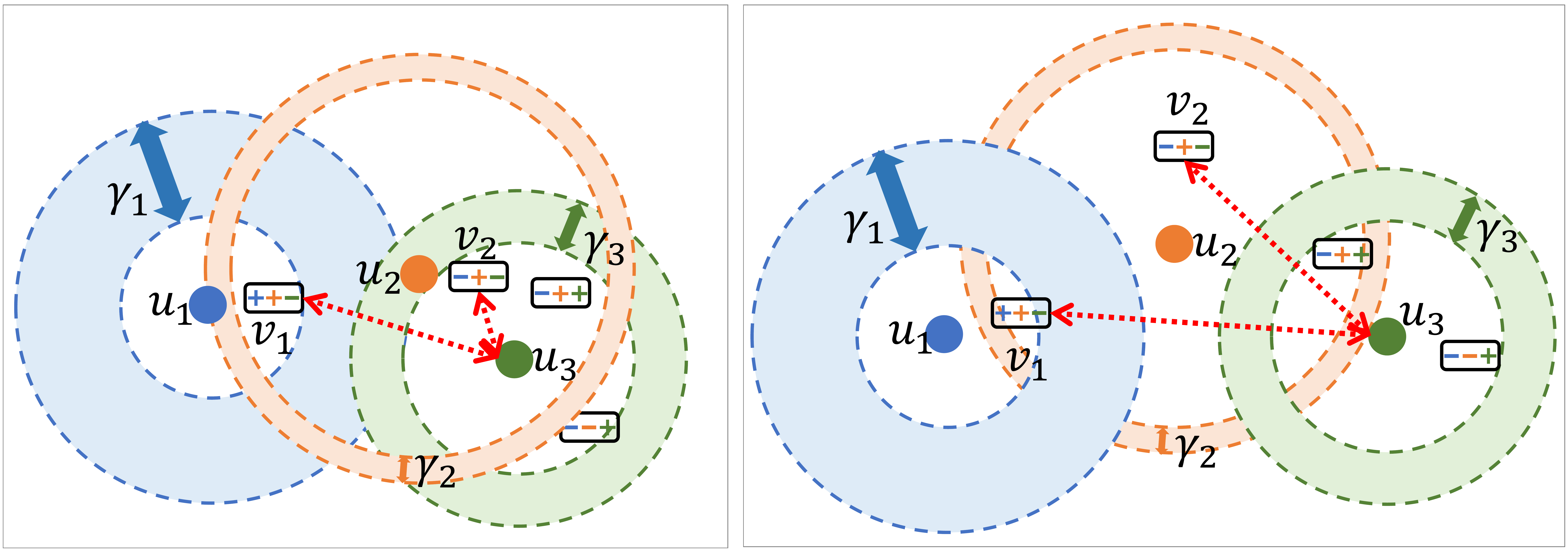}
    \label{fig:p_margin}
    }
    \caption{An illustration of single space vs.~multi-space recommendation. We highlight three users ($u_1$, $u_2$, $u_3$) and two items ($v_1$, $v_2$). ``$-+-$" of $v_1$ means user $u_2$ interacted with $v_1$ while user $u_1$ and $u_3$ did not; similarly, ``$--+$" of $v_2$ means user $u_3$ interacted with $v_2$ while user $u_1$ and $u_2$ did not.}
    \label{fig:single v.s. multi}
\end{figure*}

\subsection{Overall framework}
\label{sec:mar:a}
We use $u$ and $v$ to denote users and items, and use $\mbu \in \mathbb{R}^{D \times 1}$ and $\mbv \in \mathbb{R}^{D \times 1}$ to denote user and item embeddings, where $ D $ is the dimension of the embedding space.
We consider recommendation based on the implicit feedback matrix $\mathbf{X}$, where $ X_{uv} = 1$ corresponds to a positive sample $(\mbu, \mbv_p)$, where user $u$ interacted with item $v_p$, and $X_{uv}=0$ corresponds to a negative sample $(\mbu,\mbv_q)$, where the interaction between $u$ and $v_q$ is missing. 

To capture the multi-facet user preference and item property, a single \textit{universal embedding} $\mbu$ for user $u$ (universal item embedding $\mbv$) is projected into multiple spaces by a series of matrices $\Phi=\{\phi^k \in \mathbb{R}^{D \times D}\}_{k=1}^K$, which constitues the \textit{facet-specific embedding} $\mbU^f \triangleq\left(\boldsymbol{u}^{1}, \boldsymbol{u}^{2}, \ldots, \boldsymbol{u}^{K}\right)^{\top} \in \mathbb{R}^{K \times D}$ for user $u$ ($\mbV^f$ similarly defined for item $v$ with projection matrices $\Psi$). $K$ is the predefined number of facets, whose setting will be discussed in the experiments.
Then we compute the facet-specific similarity $\{g^k(\mbu^k, \mbv^k)\}_{k=1}^K$ between user $u$ and item $v$ in all facet spaces, which is summed up as the final similarity $g(\mbu, \mbv)$ through a learnable weight vector $\Theta \in \mathbb{R}^{K\times 1}$.

\subsection{Cross-facet similarity measurement} 
\label{sec:mar:b}
We observe that different behaviors of users come from the expression of their different preferences. \carl{Merging these different facets into a single-space representation with high dimension cannot resolve the potential conflicts.} As shown in Figure \ref{fig:multi}, users may like movies of different categories at the same time, an movies themselves may belong to multiple categories simultaneously. \carl{In a single metric space, conflicts can arise because movies from different categories are unlikely to be close, and users cannot be close to movies of different categories at the same time.}

Inspired by the ambiguity embedding method derived from word ambiguity in language modeling \cite{huang2012improving} and the argument about single embedding being less generalizable from the statistical, computational and representational point of view \cite{dietterich2000ensemble}, we propose to project users and items into multiple embedding spaces \carl{and provide a mechanism for resolving their potential conflicts}. In this way, the user ${u}$'s facet-specific embedding $\boldsymbol{U}^{f}$ can approach items with different perspectives regarding different facets at the same time. The recommended items are then calculated according to the similarity across the multi-facet user preferences. 

However, existing similarity measurements are based on single metric spaces without the consideration of multi-facet embedding. 
To connect multiple spaces and integrate the user's multi-facet representations, we propose a monotonic \textit{cross-facet similarity measurement}, which computes an overall similarity score for each user-item pair. 

Instead of calculating user-item similarity through the learned embedding directly, we design the measurement of cross-facet similarity as a three-step process: 
\begin{enumerate}
    \item The projection of universal user and item embeddings $\mbu$ and $\mbv$ into the multiple facet-specific embeddings $\mbU^f$ and $\mbV^f$ with learnable projection matrices $\Phi$ and $\Psi$.
    \item The computation of user-item similarity in each facet-specific space $g^k(\mbu^k, \mbv^k)$.
    \item The aggregation of multiple facet-specific similarities into an overall cross-facet similarity $g(\mbu, \mbv)$.
\end{enumerate}
Hence, given a user ${u}$, we first compute the facet-specific embedding as
\begin{equation}
\boldsymbol{U}^f = \left(\begin{array}{c} \boldsymbol{u}^{1\top} \\ \boldsymbol{u}^{2\top} \\ \vdots \\  \boldsymbol{u}^{K\top}  \end{array}\right) =\left(\begin{array}{c}\mbu^{\top} \phi^{1} \\ \mbu^{\top} \phi^{2} \\ \vdots \\ \mbu^{\top} \phi^{K}\end{array}\right),
\end{equation}
where $\mbu \in \mathbb{R}^{D \times 1}$ represents the user's universal preference. $\mbU^f\in \mathbb{R}^{K\times D}$ represents all facet-specific user embeddings with each $\boldsymbol{u}^{k} \in \mathbb{R}^{D \times 1}$ represents the user's specific preference towards the $k$-th facet. Each $\Phi_k$ is a $\mathbb{R}^{D \times D}$ user-facet projection matrix. The same modeling is applied to each item $v$ as
\begin{equation}
\boldsymbol{V}^f = \left(\begin{array}{c} \boldsymbol{v}^{1\top} \\ \boldsymbol{v}^{2\top} \\ \vdots \\  \boldsymbol{v}^{K\top}  \end{array}\right) =\left(\begin{array}{c}\mbv^{\top} \psi^{1} \\ \mbv^{\top} \psi^{2} \\ \vdots \\ \mbv^{\top} \psi^{K}\end{array}\right),
\end{equation}
where $\mbv\in\mathbb{R}^{D\times1}$ represents the item's universal property. $\mbV^f\in\mathbb{R}^{K\times D}$ represents all facet-specific item embeddings, and each $\Psi_k$ is a $\mathbb{R}^{D \times D}$ item-facet projection matrix. 

The usage of user-/item-specific universal embeddings and shared facet-specific projections allows the reduction of parameter space and sharing of certain cross-facet information (\eg, user activity and item popularity).

Thereby, a recommended item in the $k$-th facet is obtained according to the ranking of facet-specific similarity $g^k$ as follows
\begin{equation}
g^k\left(\boldsymbol{u}^{k},  \boldsymbol{v}^{k}\right) = -\|\mbu^k-\mbv^k\|^2,
\label{equ:euclid}
\end{equation}
where $\boldsymbol{u}^{k} \in \mathbb{R}^{D \times 1}$ and $\boldsymbol{v}^{k} \in \mathbb{R}^{D \times 1}$ are the facet-specific user and item embeddings to be learned by the model. 

Finally, to jointly capture user-item interactions in different facet-specific spaces, we compute the cross-facet similarity between user $u$ and item $v$ as
\begin{equation}
\begin{aligned}
g(\mbu, \mbv) &= \sum_{k=1}^{K} \theta^k_u  \left( g^k(\mbu^k, \mbv^k )\right) = -\sum_{k=1}^K \theta^k_u ||\mbu^k- \mbv^k||^2\\
&= -\sum_{k=1}^K \theta^k_u ||\mbu^{\top} \phi_k- \mbv^{\top} \psi_k||^2,
\label{equ:sim}
\end{aligned}
\end{equation}
where the $\Theta_u= [\theta_u^1, \ldots, \theta_u^K]$ is the vector of learnable user-specific facet weight, which is normalized to one through softmax. We devise a $\Theta_u \in \mathbb{R}^{K\times 1}$ for each user $u$, which learns to capture the importance of different facets towards the preference of each user.

\subsection{Multi-facet optimization objectives} 
\label{sec:mar:c}
With the cross-facet similarity defined, we follow the common practice of existing research on recommendation to learn the model parameters by optimizing the large margin nearest neighbor objective (LMNN) \cite{hsieh2017collaborative,park2018collaborative} as follows
\begin{equation}
\mathcal{L}_{lmnn}=\sum_{(\mbu, \mbv_p) \in \mathcal{I}} \sum_{(\mbu, \mbv_q) \notin \mathcal{I}} \left[m-g(\mbu, \mbv_p)^{2}+g(\mbu, \mbv_q)^{2}\right]_{+},
\label{equ:triplet}
\end{equation}
where $\mathcal{I}$ is the set of positive user-item pairs derived from the implicit feedback data  ($\mathbf{X}$).
Given a positive training tuple ($\mbu, \mbv_p$), the idea of LMNN is to maximize the similarity $g(\mbu, \mbv_p)$, while minimizing the similarity $g(\mbu, \mbv_q)$, where $(\mbu, \mbv_q)$ is a negative tuple, often randomly sampled from all negative tuples of $u$. $m$ is a margin hyperparameter, which is used to enforce the difference between $g(\mbu, \mbv_p)$ and $g(\mbu, \mbv_q)$. However, Eq.~\ref{equ:triplet} does not care about our embedding of users and items into mutiple facet-specific spaces, and directly optimizing it does not ensure the diversity of facet-specific spaces, thus not fully utilizing their expressive power and leading to suboptimal recommendations.

\carl{Figure \ref{fig:single v.s. multi} illustrates the scenarios of single-space versus multi-space recommendation. Besides having multiple embedding spaces, we hope to properly arrange the spaces by “spreading out” the facets, that is, we encourage them to be sufficiently diverse among themselves. In this way, we reduce the dimension redundancy, and enforce each space to capture a latent facet of user preference and item property, instead of overfitting the possible noises in a single space.}

To achieve this goal, we devise a facet-separating loss inspired by the difference loss in \cite{bousmalis2016domain,sun2020circle} to reduce the redundancy of different spaces as follows
\begin{equation}
\mathcal{L}_{facet}=\frac{1}{\alpha} \sum_{i, j} \log \left(1+ e^{ \left(-\alpha(\left\|\mbu^i - \mbu^j\right\|^{2} + \left\|\mbv^i - \mbv^j\right\|^{2} ) \right)}\right),
\label{equ:lf}
\end{equation}
where $\alpha$ is a scale hyperparameter which we empirically set to 0.1 by default in our experiments. \carl{Since the loss encourages orthogonality among different spaces, conflicting similarities among users and items can be separated into multiple spaces.}

Besides using Eq.~\ref{equ:lf} to encourage diverse facet-specific spaces from a learning perspective, we further design two techniques of \textit{adaptive margin} and \textit{explorative sampling} to explicitly improve the model's ability of capturing the multi-facet user preference and item property from limited implicit feedback data.

Recall our main objective of Eq.~\ref{equ:triplet}, where $m$ is margin hyperparameter fixed for all users and items. We find it largely limits the flexibility of arranging the users and items differently in the facet-specific spaces, thus hindering the modeling of users' diverse preferences towards different items.
Particularly, as shown in Figure \ref{fig:triplet}, based on the preferences of users $u_1$ and $u_2$ towards items $v_1$ and $v_2$, a fixed margin $m$ makes it pretty impossible to arrange the metric space in different ways, and in consequence it is hard to distinguish the preference of user $u_3$ towards $v_1$ and $v_2$. However, if we allow users to have different personalized margins like the $\gamma$'s in Figure \ref{fig:p_margin}, the model has more freedom to arrange the metrics spaces, which allows the multiple facet-specific spaces to capture user preference and item property from different perspectives.

In real-world recommendation systems, we associate the margin $m$ in Eq.~\ref{equ:triplet} with a clear physical meaning, which is related to the concept of \textit{adoption} \cite{komiak2006effects}. Specifically, users have different levels of adoption, meaning some users are more likely to adopt new things while others are not. 
Based on this intuition about the negative correlation between users' adoption levels and personalized margins, we propose to directly compute users' adoption level from the implicit feedback data and adaptively set their personalized margins as follows
\begin{equation}
\gamma_u = 1 - \frac{\sum_{v \in \mathcal{V}_u} |\mathcal{U}_v|}{N}, 
\label{equ:gamma}
\end{equation}
where $\mathcal{V}_u$ denotes the set of items that user $u$ interacts with and $\mathcal{U}_v$ denotes the set of users that item $v$ interacts with. $N$ denotes the number of users, so we have $\gamma_u \in [0, 1]$. The idea behind Eq.~\ref{equ:gamma} is to leverage the two-hop neighbors of $u$ on the bipartite user-item graph to characterize the adoption level of $u$, \ie, the more different two-hop neighbors $u$ has, the more diverse the preference of $u$ is, and thus the more likely $u$ is to adopt new things. With such adaptive margins, we rewrite Eq.~\ref{equ:triplet} as follows
\begin{equation}
    \mathcal{L}_{push}=\sum_{(\mbu, \mbv_p) \in \mathcal{I}} \sum_{(\mbu, \mbv_q) \notin \mathcal{I}}[\gamma_u-g(\mbu, \mbv_p)+g(\mbu, \mbv_q)]_{+}. \\
\label{equ:push}
\end{equation}

By incorporating adaptive margins, our model can better improve and harvest the expressive power of multiple metric spaces. In particular, for the undistinguished pairs of $(u_3, v_1)$ and $(u_3, v_2)$ in Figure \ref{fig:triplet}, our model can learn to distinguish them by properly combining the multiple different spaces of \ref{fig:p_margin} as long as they are distinguishable in one of the spaces. 

Existing works have noticed that it is impossible to simultaneously ``pull'' users and items in all positive pairs close in a single metric space \cite{hsieh2017collaborative, zhang2018metric}. However, this is no longer an issue in our multi-space setting, where all positive pairs of users and items can be close simultaneously in different spaces. To further leverage our multi-space framework and better model the reality, we enforce an additional absolute ``pulling'' objective based on the $\mathcal{L}_2$-norm pointwise regularization function as follows to complement the relative ``pushing'' objective in Eq.~\ref{equ:push} by drawing positive pairs of users and items closer in across different facet-specific spaces
\begin{equation}
\mathcal{L}_{pull}=\sum_{(\mbu, \mbv_p) \in \mathcal{I}}-g(\mbu, \mbv_p).
\label{equ:pull}
\end{equation}

Since it is impossible to traverse the cubic number of triplets in Eq.~\ref{equ:push}, sampling is widely used in existing recommendation systems. However, compared with other tasks like classification where uniform random sampling is often directly adopted, recommendation with implicit feedback is more prone to the imbalance of positive and negative samples and the ambiguity of negative samples.
The situation is even harder in our multi-facet recommendation system since the sampling has to be helpful in distinguishing all multiple facet-specific preferences instead of only the universal preference. 
To this end, we propose to bias the sampling towards more active users, to leverage their richer feedback data for the more accurate learning of different facet-specific metric spaces. 
To ensure the active users who have interacted with many items will be sampled with a high probability, we get inspired by \cite{tran2019improving, yin2017sptf} and formulate the probability function of our sampling process as follows
\begin{equation}
Pr(u) = \frac{freq(u)^{\beta}}{\sum_{u'\in \mathcal{U}} freq(u')^{\beta}},
\label{equ:nega}
\end{equation}
where $ freq(u) $ is the interaction frequency of user $u$, \ie, the number of items $u$ has interacted with. $\mathcal{U}$ is the set of all users. $ {\beta} $ is a smoothing hyperparameter, which controls the bias towards active users. We empirically set $\beta$ to 0.8 by default in our experiments.  

Finally, to avoid the problem of overfitting, we apply the widely used Euclidean sphere constraints \cite{zhang2018metric} on all facet-specific user and item embeddings, which leads to our final proposed objective function as follows
\begin{equation}
\begin{array}{l}
{\min _{\mbu_*, \mbv_*}} \quad \mathcal{L}_{push} + \lambda_{pull} \mathcal{L}_{pull} + \lambda_{facet} \mathcal{L}_{facet} \\ 
{\text { s.t. } \quad \forall k \; \left\|\mbu_*^k\right\|^{2} \leq 1 \text { and }\left\|\mbv_*^k\right\|^{2} \leq 1}.
\label{equ:loss1}
\end{array}
\end{equation}

\section{MAR with Spherical \carl{Optimization}} 
\label{sec:mars}
Our proposed MAR framework essentially projects users and items to multiple Euclidean spaces with shared projection matrices $\Phi$ and $\Psi$, and integrate them with user-specific softmax weights $\Theta$.
In Eq.~\ref{equ:loss1}, existing metric learning recommendation systems \cite{zhang2018metric} require all user and item embeddings to lie in a unit ball to avoid overfitting. However, as explored in \cite{ranjan2018crystal}, the allowance of different norms can make the model weak with difficult or extreme samples. 
In recommendation systems based on the main objective of Eq.~\ref{equ:triplet}, if we allow the user and item embeddings to have different norms (\ie, $||\mbu||^2<1$ and $||\mbv||^2<1$), the model will learn to put difficult users and items (those with insufficient or slightly contradictory training data) on the surface of the sphere (\ie, $||\mbu||^2\sim1$ or $||\mbv||^2\sim1$), so as to easily reduce the objective of Eq.~\ref{equ:loss1}, while it will only try to differentiate the easy users and items by properly arranging them inside the sphere (\ie, $||\mbu||^2<<1$ or $||\mbv||^2<<1$). In this way, the model is weak and does not generalize well due to such ``lazy'' behaviors that waste its learning capacity. 

The problem is especially concerning in our multi-facet recommendation framework, where the key is to fully exploit the learning capacity brought by multiple facet-specific embedding spaces. In particular, if we allow the user and item embeddings to have different norms, the model can learn to put its weight on all difficult users and items in a few particular facet-specific spaces where they all lie on the surfaces of the spheres, while only using the remaining spaces to model the easy users and items, thus effectively reducing the loss in Eq.~\ref{equ:loss1} without really differentiating the difficult users and items. 
In light of this, we propose a spherical optimization strategy for MAR, which enforces all facet-specific user and item embeddings to exactly lie on the surfaces of unit spheres in the different facet-specific spaces, instead of anywhere inside the sphere. In this way, the model is forced to exploit all expressive power and learn to differentiate all users and items as well as possible.
We name our framework of Multi-fAcet Recommender networks with Spherical optimization as MARS.

\subsection{Spherical cross-facet similarity and multi-facet objectives}
\label{sec:mars:a}
Different from the Euclidean space, an effective similarity measurement in the spherical space is \textit{cosine distance} \cite{shi2019probabilistic, mettes2019hyperspherical}. To maintain the objectives we design for MAR as in Eq.~\ref{equ:loss1}, we replace the similarity measure from the negative Euclidean distance to cosine similarity. The detailed differences in the objectives are described in the following.

First, we rewrite the facet-separating loss in Eq.~\ref{equ:lf} to reflect the spherical constraints on all facet-specific embeddings as follows
\begin{equation}
\mathcal{L}_{facet}^s=\frac{1}{\alpha} \sum_{i, j} \log \left(1+\exp \left(-\alpha\cos(\mbu^i, \mbu^j)\right)\right).
\label{equ:lf2}
\end{equation}

Then, instead of calculating the facet-specific similarity as the negative Euclidean distance as in Eq.~\ref{equ:euclid}, the way we recommend item in the spherical space requires the calculation of cosine similarity as follows
\begin{equation}
g_s^k\left(\mbu^k , \mbv^k\right)= \cos(\mbu^k,\mbv^k),
\label{equ:puv}
\end{equation}
where $||\mbu^k|| = ||\mbv^k|| = 1$, and $\cos(\mathbf{x}_1, \mathbf{x}_2)=\frac{\mbx_1^{\top}\mbx_2}{|\mbx_1||\mbx_2|}$ denotes the cosine of the angle between two vectors on the unit sphere.

Next, the cross-facet similarity of a user-item pair in spherical space is given by
\begin{equation}
\begin{aligned}
g_s(\mbu, \mbv) =  \sum_{k=1}^K \theta_u^k  g^k_s\left(\mbu^k, \mbv^k\right) = \sum_{k=1}^K \theta_u^k \cos(\mbu^k, \mbv^k).
\end{aligned}
\label{equ:sp}
\end{equation}
\begin{figure*}
    \centering
    \includegraphics[width=\linewidth]{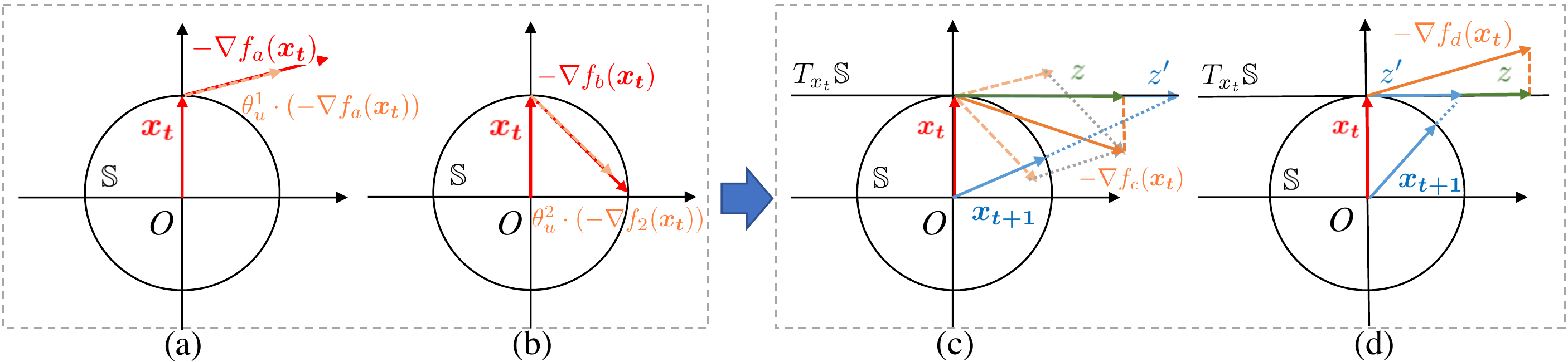}
    \caption{Illustration of calibrated Riemannian gradient descent for the optimization of MARS in the spherical space. (a) and (b) show different gradients in single spaces. After weighted-combining the single-space gradients, (c) shows the multi-space gradient. In (c) and (d), $\boldsymbol{z} = \operatorname{grad} f(\mbx_t)$ is the original Riemannian gradient and $\boldsymbol{z}'=1+\frac{\mbx_t^{\top} \nabla f(\mbx_t)}{\|\nabla f(\mbx_t)\|}$ is the calibrated one, which takes the angular distances into account during parameter updates.}
    \label{fig:RSGD}
\end{figure*}

Based on Eq.~\ref{equ:sp}, $\mathcal{L}_{push}^s$ on spherical space is formulated as follows
\begin{equation}
\begin{aligned} 
\mathcal{L}_{push}^s &= \sum_{(\mbu, \mbv_p) \in \mathcal{I}} \sum_{(\mbu, \mbv_q) \notin \mathcal{I}} [\gamma_u - g_s(\mbu, \mbv_p)^2 + g_s(\mbu, \mbv_q)^2]_+\\
&=\sum_{(\mbu, \mbv_p) \in \mathcal{I}} \sum_{(\mbu, \mbv_q) \notin \mathcal{I}} [\gamma_u - \sum_{k=1}^{K} \theta_u^k \cos (\mbu^{\top} \Phi^k, \mbv_p^{\top} \Psi^k) \\
&+ \sum_{k=1}^{K} \theta_u^k \cos (\mbu^{\top} \Phi^k,\mbv_q^{\top} \Psi^k)]_+.
\end{aligned}
\end{equation}

Similarly, we rewrite $\mathcal{L}_{pull}$ as follows
\begin{equation}
    \mathcal{L}^s_{pull} 
    = -\sum_{(\mbu, \mbv_p) \in \mathcal{I}} \sum_{k=1}^{K}\theta_u^k \cos (\mbu^{\top} \Phi^k, \mbv_p^{\top} \Psi^k).
\end{equation}

Finally, we get the objective function in spherical space as
\begin{equation}
\begin{array}{l}
{\min _{\mathbf{u_*}, \mathbf{v_*}}} \quad \mathcal{L}_{push}^s + \lambda_{pull} \mathcal{L}_{pull}^s + \lambda_{facet} \mathcal{L}_{facet}^s \\ 
{\text { s.t. } \quad \forall k \; \left\|\mbu^k_*\right\|^{2} = 1 \text { and }\left\|\mbv^k_*\right\|^{2} = 1}.
\label{equ:loss}
\end{array}
\end{equation}

Note that, the direct modeling of user-item similarity with cosine similarity in the spherical space can be also interpreted in a probabilistic point of way. Suppose we model the probability of recommending an item $v$ to a user $u$ regarding the $k$-th preference/property facet with the conditional probability $p(\mbv^k|\mbu^k)$. By drawing analogy from spherical text embedding \cite{meng2019spherical} which models the target-context word interactions with $p(w_t|w_c)$, we have 
\begin{equation}
\begin{aligned}
    p(\mbv^k|\mbu^k) &\propto \operatorname{vMF}_{p}(\mbv^k ; \mbu^k, 1) \propto c_p(1)\exp(\cos(\mbv^k,\mbu^k)).
\end{aligned}
\end{equation}
Here vMF is the von Mises-Fisher distribution, which defines a probability density over the hypersphere and is parameterized by a mean vector $\mu$ and a concentration parameter $\kappa$ \cite{tian2019sosnet,banerjee2005clustering}. 
Formally, a unit random vector $\mbx \in \mathbb{S}^{p-1}$ has the $p$-variate vMF distribution $\operatorname{vMF}_{p}$($\mbx$; $\mu$, $\kappa$) if its probability dense function is $f(\mbx ; \mu, \kappa)=c_{p}(\kappa) \exp (\kappa \cos (\mbx, \mu))$, 
where $||\mu|| = 1$ is the mean direction, $\kappa \le 0$ is the concentration parameter, and $c_{p}(\kappa)$ is the normalization constant.
In the limit case when $|\mathcal{V}|\to\infty$, the analytic form of $p(\mbv^k|\mbu^k)$ follows the vMF distribution with the prior embedding as the mean direction and constant 1 as the concentration parameter, \ie, $\lim _{|V| \rightarrow \infty} p(\mbv^k |\mbu^k)=\operatorname{vMF}_{p}(\mbv^k ; \mbu^k, 1)$ \cite{meng2019spherical}. The modeling of $p(\mbv^k|\mbu^k)$ in a maximum likelihood estimation setting leads to $\log p(\mbv^k|\mbu^k)\propto \cos(\mbv^k,\mbu^k)$.

\subsection{Calibrated Riemannian optimization}
\label{sec:mars:b}
To improve the modeling of difficult users and items, we propose to constrain all facet-specific user and item embeddings to lie on the surface of a sphere, so as to avoid trivial responses to the loss.
The unit hypersphere $\mathbb{S}^{p-1} := \{\boldsymbol{x} \in \mathbb{R}^p \ | \ ||\boldsymbol{x}|| = 1\}$ is a common choice for optimization problems with spherical constraints. 
In our case, the learning of MARS is thus a constrained optimization problem:
\begin{equation}
\min _{\Omega} \mathcal{L}(\Omega) \quad \text { s.t. } \quad \forall \omega \in \Omega:\|\omega\|=1
\end{equation}
where $\Omega =  \{ \mbu^k\}_{k=1}^{K} \bigcup \{ \mbv^k\}_{k=1}^{K}$ is the set of user and item embeddings in all facet-specific spaces. 
The Euclidean gradients provide the update directions in a non-curvature space. However, the parameters in the proposed model must be updated on a surface with constant positive curvature, where the Euclidean space optimization methods such as SGD cannot work. Therefore, we introduce calibrated Riemannian optimization for the proposed MARS framework.

Since the unit sphere is a Riemannian manifold, we can optimize our objectives with Riemannian SGD \cite{bonnabel2013stochastic}. Specifically, the parameters are updated by 
\begin{equation}
\mbx_{t+1} = {\rm exp}_{\mbx_t} (-\eta_t \operatorname{grad} f(\mbx_t))
\label{eq:rsgd}
\end{equation}
where $\eta_t$ denotes the learning rate and $\operatorname{grad} f(\mbx_t) \in T_{\mbx}\mathbb{S}^{p-1}$ is the Riemannian gradient of a differentiable function $f: \mathbb{S}^{p-1} \to \mathbb{R}$. More details of Eq.~\ref{eq:rsgd} can be found in \cite{skopek2019mixed}.

Figure \ref{fig:RSGD} gives an illustration of our calibrated Riemannian gradient descent framework. The callibration is done in two perspectives. 
First, in the comparison among (a), (b) and (c), the Euclidean gradient in multi-space is a combination of multiple gradients (\ie, $\nabla f_c(\mbx_t) = \nabla f_a(\mbx_t)+\nabla f_b(\mbx_t)$). 
Second, in the comparison between (c) and (d), $\nabla f(\mbx_t)$ is projected to the tangent of $T_{\mbx_t}\mathbb{S}$, which provides the correct direction to update the parameters on the sphere, but ignores the different angles between $\nabla f(\mbx_t)$ and $\mbx_t$. As a consequence, $\nabla f_c(\mbx_t)$ and $\nabla f_d(\mbx_t)$ will lead to the same $\mbx_{t+1}$, which is not ideal because we care about cosine similarities and thus the different angles. To explicitly incorporate angular distance into the optimization, we get inspired by \cite{meng2019spherical} to use the cosine similarity between $\mbx_t \in \mathbb{S}^{p-1}$ and the Euclidean gradient $\nabla f(\mbx_t)$, \ie, $1+\frac{\mbx_t^{\top} \nabla f(\mbx_t)}{\|\nabla f(\mbx_t)\|}$, as a multiplier to calibrate the Riemannian gradient and update the model parameters as
\begin{equation}
\boldsymbol{x}_{t+1}=R_{\boldsymbol{x}_{t}}\left(-\eta_{t}\left(1+\frac{\boldsymbol{x}_{t}^{\top} \nabla f\left(\boldsymbol{x}_{t}\right)}{\left\|\nabla f\left(\boldsymbol{x}_{t}\right)\right\|}\right)\left(I-\boldsymbol{x}_{t} \boldsymbol{x}_{t}^{\top}\right) \nabla f\left(\boldsymbol{x}_{t}\right)\right),
\label{eq:crsgd}
\end{equation}
where $R_{\boldsymbol{x}} (\boldsymbol{z}) = \frac{\boldsymbol{x}+\boldsymbol{z}}{\|\boldsymbol{x}+\boldsymbol{z}\|}$ \cite{skopek2019mixed}. The rationale is to encourage parameters with greater cosine distance from its target direction to take a larger update step, as shown in Figure \ref{fig:RSGD}, where $z'$ in (c) is larger than that in (d). 

\carl{Note that, compared with Eq.~\ref{eq:rsgd}, our callibrated Riemannian gradient descent in Eq.~\ref{eq:crsgd} does not introduce significantly more computations. 
In our experiments, we also find the runtimes of both MAR and MARS to be in the same scale with most metric learning baselines.}

\section{Experiments}
\label{sec:exp}
In this section, we evaluate our proposed MAR and MARS frameworks focusing on the following four research questions:
\begin{itemize}
  \item\textbf{RQ1: }How do MAR and MARS perform in comparison to state-of-the-art single space recommendation methods?
  \item\textbf{RQ2: }What are the effects of different model components? 
  \item\textbf{RQ3: }How do the hyperparameters affect the recommendation performance and how to choose optimal values?
  \item\textbf{RQ4: }How do MAR and MARS improve the modeling of multiple facets of users and items? 
\end{itemize}

\subsection{Experimental setup}

\begin{table}
  \caption{Statistics of the datasets used in our experiments.}
  \centering
  \begin{tabular}{p{1.2cm}|p{1.1cm}|p{1.1cm}|p{1.7cm}|p{1.2cm}}
      \hline
      \hline
      Dataset & \# \ User & \# \ Item & \# \ Interaction & Density(\%) \\
      \hline
      Delicious & 1K & 1K & 8K & 0.61 \\
      \hline
      Lastfm & 2K & 175K & 92K & 0.28 \\
      \hline
      Ciao & 7K & 11K & 147K & 0.19 \\
      \hline
      BookX & 20K & 40K & 605K & 0.08 \\
      \hline
      \carl{ML-1M} & 6K & 4K & 1M & 4.52\\
      \hline
      \carl{ML-20M} & 62K & 27K & 17M & 1.02\\
      \hline
      \hline
  \end{tabular}
\label{tab:datasets-stat}
\end{table}

\subsubsection{Datasets}
In order to comprehensively verify the effectiveness of compared methods, we use \carl{six} real-world datasets from different application domains with different sizes and interaction density, \ie, Delicious\textsuperscript{\ref{data}}, Lastfm\footnote{http://millionsongdataset.com/lastfm/},  Ciao\footnote{https://github.com/pcy1302/TransCF/tree/master/data\label{data}}, BookX\textsuperscript{\ref{data}}, \carl{ML-1M\footnote{https://grouplens.org/datasets/movielens/\label{ml}} and ML-20M\textsuperscript{\ref{ml}}}. These datasets have been widely adopted in previous literature \cite{hsieh2017collaborative,park2018collaborative,zhang2018metric}, and their statistics are summarized in Table \ref{tab:datasets-stat}.  

\subsubsection{Evaluation protocols}
We follow the standard evaluation protocols as in \cite{park2018collaborative}. In particular, we adopt the leave-one-out evaluation, \ie, the testing set comprises the last item of all users. If there are no timestamps available in the dataset (\eg, Delicious), the test sample is randomly selected. One item for each user is also sampled to form the development set. Since it is too time-consuming to rank all items for every user, we randomly sample 100 items that have no interactions with the target user and rank the test item with respect to these 100 items, which is a standard method for recommendation evaluation \cite{he2016ups,he2016vbpr,he2017translation,he2017neural,tay2018latent,bayer2017generic,he2016fast,rendle2009bpr,xue2017deep}. Since our problem is essentially formulated as learning-to-rank, we judge the performance of compared models based on the standard metrics in information retrieval and recommender systems: hit ratio (HR@10, HR@20) \cite{he2017neural} and normalized discounted cumulative gain (nDCG@10, nDCG@20) \cite{jarvelin2002cumulated}. Intuitively, the HR metric simply considers whether the ground truth is ranked amongst the top $N$ items while the nDCG metric is a position-aware ranking metric.

\begin{table*}
  \caption{Experimental results on \carl{six} benchmark datasets. Best performances are in boldface and the second runners are underlined. MARS achieves the best performance on all datasets. \emph{Imp1.}~denotes the relative improvements of MAR over the best baselines and \emph{Imp2.}~denotes those of MARS over the best baselines.}
  \centering
  \begin{tabular}{c|c|cccccccc|cc|cc}
  \hline
  \hline
     Dataset& Metric & BPR & NMF & NeuMF & CML & MetricF & TransCF & LRML & SML & \textbf{MAR} & \textbf{MARS} & \emph{Imp1.} & \emph{Imp2.}\\
    \hline
    \multirow{4}{*}{\rotatebox{90}{Delicious}}
    & HR@10 & 0.1981 & 0.2031 & 0.1164 & 0.2470 & 0.2137 & \underline{0.2586} & 0.2124 & 0.2071 &  0.3298 & \textbf{0.3311} & 27.53\% & 28.04\%  \\
    & HR@20 & 0.3177 & 0.3100 & 0.2171 & 0.3649 & 0.3245 & \underline{0.3786} & 0.3285 & 0.3285 & 0.4697 & \textbf{0.4842} & 24.06\% & 27.89\%\\
    & nDCG@10 & 0.1122 & 0.1113 & 0.0558 & 0.1389 & 0.1128 & \underline{0.1475} & 0.1199 & 0.1114 & 0.1828 &\textbf{0.1869} & 23.93\% & 26.71\% \\
    & nDCG@20 & 0.1418 & 0.1383 & 0.0789 & 0.1678 & 0.1391 & \underline{0.1781} & 0.1469 &0.1406 & 0.2144 &\textbf{0.2234} & 20.38\% & 25.44\% \\
    
    \hline    
    \multirow{4}{*}{\rotatebox{90}{Lastfm}}
    & HR@10 & 0.2073 & 0.1965 & 0.1489 & 0.1975 & 0.2045 & 0.2211 & 0.2171 & \underline{0.2280} & 0.2555 & \textbf{0.2818} & 12.06\% & 23.62\%\\
    & HR@20 & 0.2488 & 0.2279 & 0.1697 & 0.2455 & 0.2412 & 0.2749 & 0.2699 & \underline{0.2834} & 0.3183& \textbf{0.3425} & 12.31\% & 20.86\%\\
    & nDCG@10 & 0.1358 & 0.1332 & 0.0961 & 0.1203 & 0.1307 & 0.1465 & 0.1438  & \underline{0.1510} & 0.1755 & \textbf{0.1882} & 16.23\% & 24.64\%\\
    & nDCG@20 & 0.1480 & 0.1425 & 0.1009 & 0.1331 & 0.1418 & 0.1621 & 0.1591 & \underline{0.1671}  & 0.1892 & \textbf{0.2041} & 13.22\% & 22.14\%\\
    
    \hline
    \multirow{4}{*}{\rotatebox{90}{Ciao}}
    & HR@10 & 0.1569 & 0.1549 & 0.1535 & 0.2085 & 0.1722 & 0.2292 & 0.2122 & \underline{0.2307} & 0.2480 &\textbf{0.3393} & 7.50\% & 47.07\% \\
    & HR@20 & 0.2811 & 0.2741 & 0.2788 & 0.3337 & 0.3012 & \underline{0.3740} & 0.3345 & 0.3494 & 0.3985 &\textbf{0.5097} & 6.55\% & 36.28\% \\
    & nDCG@10 & 0.0751 & 0.0798 & 0.0741 & 0.1053 & 0.0863 & 0.1167 & 0.1102 & \underline{0.1208} & 0.1299 &\textbf{0.1776} & 7.53\% & 47.02\% \\
    & nDCG@20 & 0.1063 & 0.1097 & 0.1040 & 0.1358 & 0.1180 & \underline{0.1525} & 0.1406 & 0.1509 & 0.1673 &\textbf{0.2202} & 9.70\%  & 44.39\% \\
    \hline
    \multirow{4}{*}{\rotatebox{90}{BookX}}
    & HR@10 & 0.2425 & 0.1981 & 0.2286 & 0.2885 & 0.2418 & 0.3329 & 0.3168 & \underline{0.3367} & 0.3697 &\textbf{0.4162} & 9.80\% & 23.61\% \\
    & HR@20 & 0.3761 & 0.3195 & 0.3747 & 0.4053 & 0.3742 & \underline{0.4744} & 0.4463 & 0.4710  & 0.5219 &\textbf{0.5851} & 10.01\% &  23.33\%\\
    & nDCG@10 & 0.1250 & 0.1041 & 0.1158 & 0.1663 & 0.1358 & 0.1865  & 0.1847  & \underline{0.2032} &0.2106 & \textbf{0.2349} & 3.64\% & 15.60\% \\
    & nDCG@20 & 0.1585 & 0.1344 & 0.1482 & 0.1956 & 0.1689 & 0.2221 & 0.2171 & \underline{0.2352} & 0.2484 &\textbf{0.2772} & 5.61\% & 17.86\% \\
    \hline
    \carl{\multirow{4}{*}{\rotatebox{90}{ML-1M}}}
    & HR@10 & 0.7237&0.7029&0.6861&0.7216&0.7198&0.7233&\underline{0.7397}&0.7126&0.7523&\textbf{0.7603}&1.70\% &2.78\% \\
    & HR@20 & 0.854&0.8294&0.8096&0.8515&0.8494&0.8535&\underline{0.8728}&0.8409&0.8832&\textbf{0.8983}&1.19\%&2.92\%\\
    & nDCG@10 & 0.5333&0.518&0.5065&0.5413&0.5304&0.5330&\underline{0.5461}&0.5345&0.5667&\textbf{0.5897}&3.77\%&7.98\%\\
    & nDCG@20 & 0.6164&0.5987&0.5854&0.6256&0.6131&0.6160&\underline{0.6311}&0.6178&0.6681&\textbf{0.6801}&5.86\%&7.76\% \\
    \hline
    
    \carl{\multirow{4}{*}{\rotatebox{90}{ML-20M}}}
    & HR@10 & 0.8220 &0.8175&0.7944&0.8457&0.8192&\underline{0.8504}&0.8132&0.8216&0.8678&\textbf{0.8799}&2.05\%&3.47\% \\
    & HR@20 & 0.9005&0.8842&0.8725&0.9106&0.9107&\underline{0.9113}&0.8855&0.9025&0.9311&\textbf{0.9466}&2.17\%&3.87\%\\
    & nDCG@10 & 0.6346&0.593&0.5815&0.6256&0.6158&0.6239&\underline{0.6377}&0.6257&0.6632&\textbf{0.6714}&4.00\%&5.28\% \\
    & nDCG@20 & 0.6566&0.6214&0.6291&0.6594&0.6568&0.6633&\underline{0.6705}&0.6529&0.6829&\textbf{0.6925}&1.85\%&3.28\% \\
    \hline
    \hline
  \end{tabular}
    \label{tab:exp-all}
\end{table*}

\begin{table}[]
    \centering
    \caption{\carl{Performance under different settings of embedding dimension.}}
    \begin{tabular}{c|cccc|cc}
        \hline
        \hline
         &HR@10&HR@20&nDCG@10&nDCG@20&$d$&$k$\\
         \hline
\hline
\multirow{4}{*}{\rotatebox{90}{TransCF}}
&0.2251&0.3587&0.1161&0.1479&128&1\\
&0.2287&0.3662&0.1164&0.1495&256&1\\
&\textbf{0.2292}&\textbf{0.3740}&\textbf{0.1167}&\textbf{0.1525}&\textbf{512}&\textbf{1}\\
&0.2224&0.3566&0.1155&0.1466&1024&1\\
\hline
\multirow{4}{*}{\rotatebox{90}{SML}}&0.2297&0.3438&0.1162&0.1437&128&1\\
&0.2302&0.3488&0.1178&0.1499&256&1\\
&\textbf{0.2307}&\textbf{0.3494}&\textbf{0.1208}&\textbf{0.1509}&\textbf{512}&\textbf{1}\\
&0.2299&0.3481&0.1183&0.1473&1024&1\\
\hline
\multirow{4}{*}{\rotatebox{90}{MARS}}
&0.3340&0.4999&0.1737&0.2158&32&4\\
&0.3371&0.5063&0.1749&0.2163&64&4\\
&0.3386&0.5027&0.1754&0.2167&128&4\\
&\textbf{0.3393}&\textbf{0.5097}&\textbf{0.1776}&\textbf{0.2202}&\textbf{256}&\textbf{4}\\
        \hline
        \hline
    \end{tabular}
    \label{tab:dim}
\end{table}

\subsubsection{Baselines}
We adopt the following representative state-of-the-art methods as baselines for performance comparison:
\begin{itemize}[leftmargin=10pt]
  \item\textbf{BPR}\cite{rendle2009bpr}: The Bayesian personalized ranking model is a popular method for Top-N recommendation. We adopt matrix factorization as the prediction component.
  \item\textbf{NMF}\cite{lee1999learning}: Non-negative matrix factorization (NMF) is a classic model that learns latent factors from interaction data. We include NMF as one of the baselines because we apply it to initialize the multiple facets of users and items. The number of latent factors is set to the same as the number of metric spaces in our proposed models.
  \item\textbf{NeuMF}\cite{he2017neural}: NeuMF is a framework for applying neural networks to collaborative filtering. It combines multiple perceptrons and matrix factorization in its framework.
  \item\textbf{CML}\cite{hsieh2017collaborative}: Collaborative metric learning (CML) is the first model to use metric learning to solve the collaborative filtering problem of recommender systems.
  \item\textbf{MetricF}\cite{zhang2018metric}: MetricF is a metric learning method that converts user preferences into distances, which uses Euclidean distance instead of the dot product.
  \item\textbf{TransCF}\cite{park2018collaborative}: TransCF borrows the idea of translation in knowledge graph embedding to improve CML, and calculates the distance metric by learning the relationship vector between users and items.
  \item\textbf{LRML}\cite{tay2018latent}: Latent relational metric learning (LRML) employs an augmented memory module to induce a latent relation for each user-item interaction.
  \item\textbf{SML}\cite{lisymmetric}: Symmetic metric learning with learnable margins introduces a symmetrical positive item-centric metric to pull and push items via the dynamic margin.
\end{itemize}

\subsubsection{Implementation Details}
We implement MAR and MARS with Pytorch\footnote{https://pytorch.org/}, which will be published upon the acceptance of this work.
Implementations of the compared baselines are either from open-source project or the original authors (BPR/MetricF/CML\footnote{https://github.com/cheungdaven/DeepRec}, NMF\footnote{https://github.com/ninghaohello/Polysemous-Network-Embedding/}, NeuMF\footnote{https://github.com/hexiangnan/neural\_collaborative\_filtering}, TransCF\footnote{https://github.com/pcy1302/TransCF}, LRML\footnote{https://github.com/vanzytay/WWW2018\_LRML} and SML\footnote{https://github.com/MingmingLie/SML}). We optimize MAR with standard SGD and MARS with Riemannian SGD as introduced in Section \ref{sec:mars:b}. We tune all hyperparameters on the validation set through grid search, in particular, $K$ in $[1, 6]$, learning rate in $\{0.0005, 0.001, 0.005, 0.01, 0.1\}$, $\lambda_{pull}$ 
and $\lambda_{facet}$ in $\{0.001, 0.01, 0.1, 1\}$ and the embedding size in $\{32, 64, 128, 256, 512, 1024\}$ for different datasets. The batch size is set to 1000. 
We also carefully tuned the hyperparameters of all baselines through cross validation as suggested in the original papers to achieve their best performance.

\subsection{Overall performance comparison (RQ1)}
We compare the recommendation results of the proposed MAR and MARS frameworks to those of the baseline models.
Table \ref{tab:exp-all} shows the HR@$K$ and nDCG@$K$ scores on all \carl{six} datasets with $K$ = $\{10, 20\}$. We have the following observations:

In general, MAR and MARS both outperform all eight baselines across all evaluation metrics on all datasets, including both dense datasets like Delicious \carl{and ML1M} and sparse datasets like Ciao and BookX. This answers RQ1, showing that our proposed multi-facet recommendation framework is capable of effective collaborative ranking. Moreover, the ranking of many baselines is fluctuating across datasets as we see the second best performance scattered among different models like TransCF and SML. Compared with the second best performance, the performance gains of MARS on the Delicious, Lastfm, Ciao, BookX, \carl{ML-1M and ML-20M} datasets range from reasonably large (\carl{2.78\% achieved with HR@10 on the ML-1M dataset}) to significantly large (47.07\% achieved with HR@10 on the Ciao dataset).

In particular, the six models based on metric learning (CML, MetricF, TransCF, LRML, SML and MAR) obviously outperform the MF–based competitors (BPR, NMF, and NeuMF), which is consistent with the results in previous work \cite{hsieh2017collaborative}. 
Moreover, we observe that our basic MAR framework already achieves the best performance compared with all single space metric learning methods, strongly indicating its effectiveness in learning and utilizing the multi-facet user preference and item property for recommendation. This is particularly evident on datasets with richer multi-facet properties, where MAR significantly outperforms other metric learning algorithms by up to 27.53\% on Delicious, 16.23\% on Lastfm, 9.70\% on Ciao,  10.01\% on BookX, \carl{5.86\% on ML-1M and 4.00\% on ML-20M}.

One step further, MARS improves over the state-of-the-art and MAR on all datasets, especially those sparser ones, where it consistently outperforms the state-of-the-art by 47.07\% and 23.61\% with HR@10 on Ciao and BookX datasets, respectively, while MAR only achieves 7.50\% and 9.80\% improvements. This strongly indicates the effectiveness of our spherical optimization strategy. 

\carl{Moreover, Table \ref{tab:dim} shows the performances of TransCF, SML and our proposed MARS with varying settings of embedding dimension $d$. The total embedding dimension of TransCF and SML is equal to $d$, while that of MARS is equal to $d\times k$. We have observed that the performances of both single-space and multi-space models slightly ascend as $d$ increases before the total embedding dimension is too large. However, the improvements brought by increased embedding dimension (such as 1.82\% on TransCF and 0.44\% on SML on HR@10) are minor compared with the gaps between the single-space models and MARS (MARS outperforms TransCF by up to 52.56\% and SML by up to 47.59\% on HR@10 under the same total dimensions), which directly supports our argument that properly using multiple embedding spaces is more effective than simply increasing the dimension of a single space. Note that, when the dimension of a single space increases to 1024, both TransCF and SML suffer from slight overfitting, while MARS can still further improve the performance.}

\begin{table*}[]
    \centering
    \caption{nDCG@10 of CML, MAR and MARS over different numbers of facet-specific spaces on \carl{four of the} datasets. \emph{Imp1.}~denotes the relative improvements of MAR over CML, \emph{Imp2.}~denotes those of MARS over CML and \emph{Imp3.}~denotes those of MARS over MAR. }
    \begin{tabular}{c ccc ccc  c  ccc ccc }
    \hline
    \hline
    \multirow{2}{*}{K spaces} & \multicolumn{6}{c}{Delicious} &
    \multirow{2}{*}{K spaces} &
    \multicolumn{6}{c}{Lastfm} \\
     & CML   & MAR  & MARS  & \emph{Imp1.} & \emph{Imp2.} & \emph{Imp3.}  & 
     & CML   & MAR  & MARS  & \emph{Imp1.} & \emph{Imp2.} & \emph{Imp3.} \\
    \hline
    K=1 & \multirow{6}{*}{0.1389} & 0.1687 & 0.1865  & 21.45\% & 34.27\% & 10.55\% & 
    K=1 & \multirow{6}{*}{0.1231}  & 0.1654 &0.1834 &37.53\% &52.42\%  &10.82\% \\
    K=2 & & \underline{0.1828} & \textbf{0.1869}  & 31.61\% &
     34.56\% & 2.24\% & 
     K=2 & &\underline{0.1755} & \textbf{0.1882} &45.89\% &56.44\%  &7.24\%\\
     
    K=3 &  & 0.1827 & 0.1846  & 31.53\% & 
    32.90\% & 1.04\% & 
     K=3& &0.1669 &0.1780 &38.76\% &47.99\% &6.65\% \\
    K=4 &  & 0.1794 & 0.1847  &29.16\% &
    32.97\% & 2.95\%  & 
     K=4 & &0.1505 &0.1778 &25.08\% &47.83\% &18.19\%\\
    K=5 &  & 0.1786 & 0.1851  & 28.58\% & 
    33.26\% & 3.64\%  & 
     K=5& &0.1434 &0.1739 &19.22\% &44.56\% &21.26\% \\
    K=6 &  & 0.1780 & 0.1856  & 28.15\% & 
    33.62\% & 4.27\% & 
     K=6& &0.1374 &0.1683 &14.22\% &39.86\% &22.45\% \\
    \hline
    \hline
    \multirow{2}{*}{K spaces} & \multicolumn{6}{c}{Ciao} &
    \multirow{2}{*}{K spaces} &
    \multicolumn{6}{c}{BookX} \\
     & CML   & MAR  & MARS  & \emph{Imp1.} & \emph{Imp2.} & \emph{Imp3.}  & 
     & CML   & MAR  & MARS  & \emph{Imp1.} & \emph{Imp2.} & \emph{Imp3.} \\
    \hline
    K=1 & \multirow{6}{*}{0.1053} & 0.1299 & 0.1667  & 23.36\% & 58.27\% & 28.30\% & 
    K=1 & \multirow{6}{*}{0.1663}  & 0.1831 &0.2307 &10.10\% &38.73\%  &26.00\% \\
    
    K=2 &  & 0.1480  & 0.1698  & 40.55\% &
    61.28\% & 14.75\% 
    & 
     K=2& &0.2002 &0.2306 &20.38\% & 38.67\%  &15.18\%\\
    
    K=3 & &0.1549 & 0.1714  & 47.10\% &
    62.77\% & 10.65\%  
    & 
     K=3& &\underline{0.2238} &0.2324 &34.58\% &39.75\% &3.84\% \\
    K=4 &  & \underline{0.1625} & \textbf{0.1776}  & 54.32\% & 
    68.66\% & 9.29\%
    & 
     K=4& &0.2176 &0.2324 &30.85\% &39.75\% &6.80\%\\
    K=5 &  & 0.1479 & 0.1682  & 40.46\% & 
    59.78\% & 13.76\%  & 
     K=5& &0.1943 & \textbf{0.2349} &16.84\% &41.25\% &20.90\% \\
    K=6 &  & 0.1346 & 0.1620  & 27.83\% & 
    53.86\% & 20.37\% & 
     K=6& &0.1806 &0.2320 &8.60\% & 39.51\% & 28.46\% \\
    \hline
    \hline
    \end{tabular}
    \label{table:ablation}
\end{table*}

\subsection{Model ablation study (RQ2)}
\label{sec:rq2}
To better understand our proposed techniques, \ie, multi-facet embedding and spherical optimization, we closely study the two frameworks of MAR and MARS.
Specifically, to study the effectiveness of multi-facet embedding, we set the classic single space metric learning method CML as the baseline and study the performance of MAR as we use different numbers of embedding spaces ($K$); to further study the effectiveness of spherical optimization, we also compare MARS to MAR in all those settings.

\begin{figure*}
\centering
  \subfigure[Varying $\lambda_{pull}$ on Delicious]{
    \includegraphics[width = 0.223\linewidth]{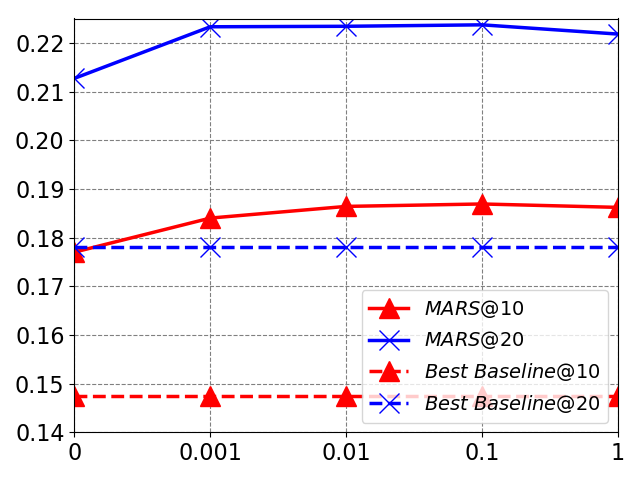}
  }
  \subfigure[Varying $\lambda_{pull}$ on Lastfm]{
    \includegraphics[width = 0.223\linewidth]{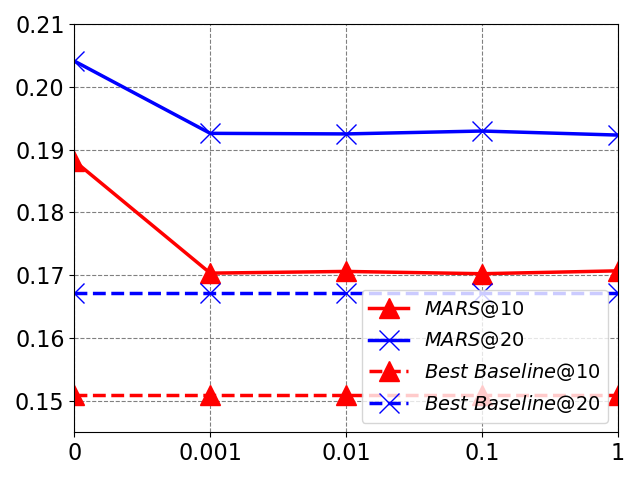}
  }
  \subfigure[Varying $\lambda_{pull}$ on Ciao]{
    \includegraphics[width = 0.223\linewidth]{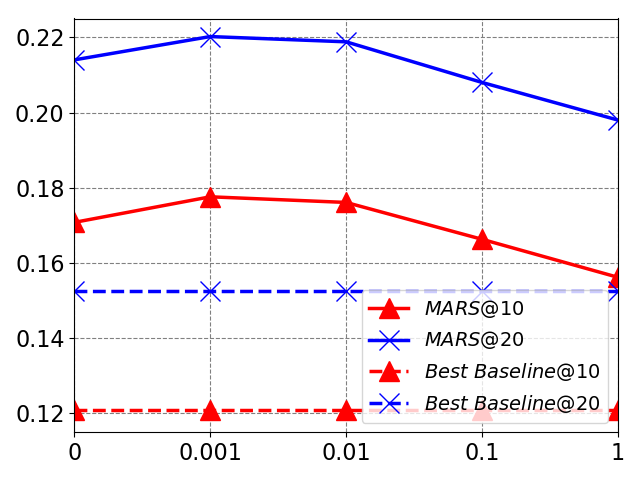}
  }
  \subfigure[Varying $\lambda_{pull}$ on BookX]{
    \includegraphics[width = 0.223\linewidth]{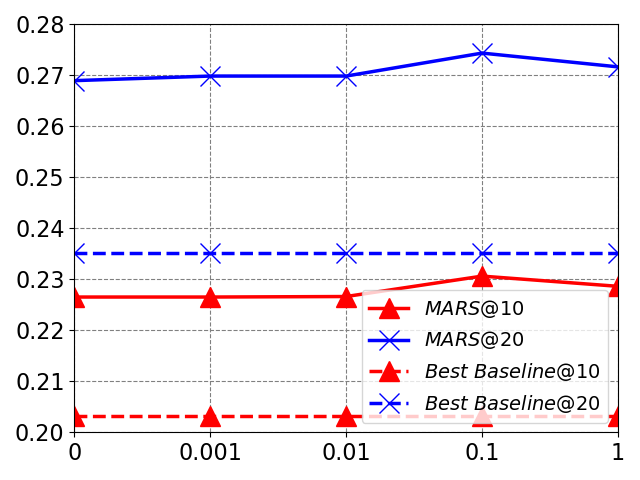}
  }
    \caption{Performance regarding nDCG of the best baseline and MARS with varying weights on ``pulling'' regularizer on \carl{four of the} datasets.}
  \label{fig:pull}
\end{figure*}

\begin{figure*}
\centering
  \subfigure[Varying $\lambda_{facet}$ on Delicious]{
    \includegraphics[width = 0.223\linewidth]{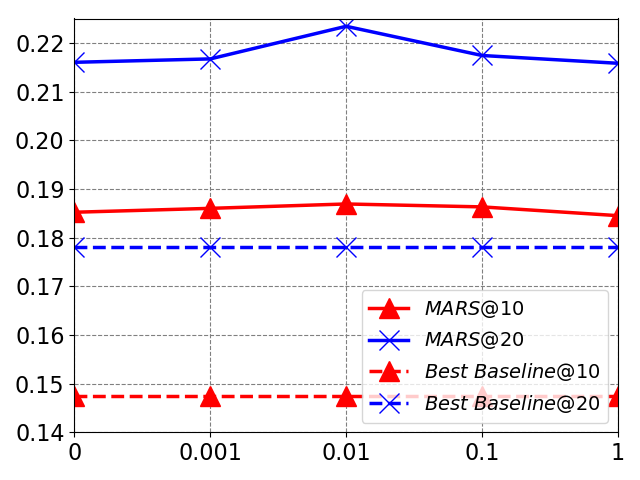}
  }
  \subfigure[Varying $\lambda_{facet}$ on Lastfm]{
    \includegraphics[width = 0.223\linewidth]{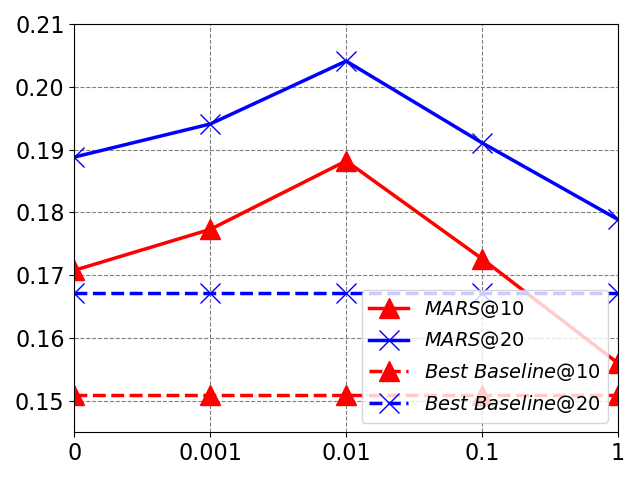}
  }
  \subfigure[Varying $\lambda_{facet}$ on Ciao]{
    \includegraphics[width = 0.223\linewidth]{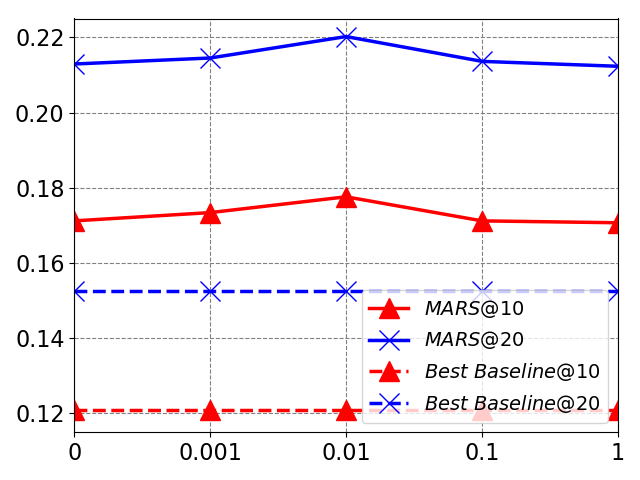}
  }
  \subfigure[Varying $\lambda_{facet}$ on BookX]{
    \includegraphics[width = 0.223\linewidth]{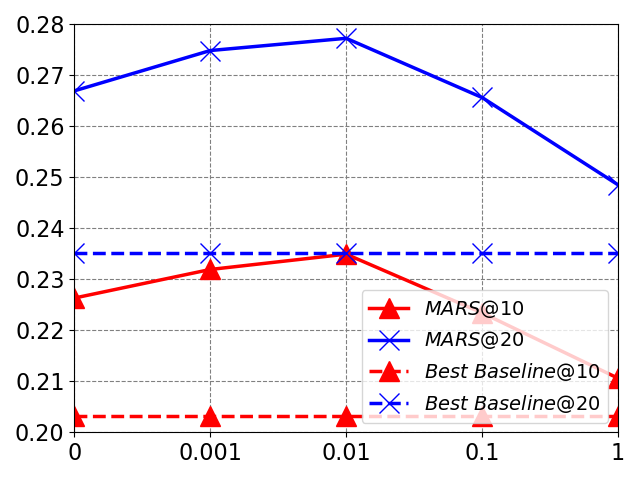}
  }
  \caption{Performance regarding nDCG of the best baseline and MARS with varying weights on facet-separating regularizer on \carl{four of the} datasets.}
  \label{fig:facet}
\end{figure*}

From Table \ref{table:ablation}, we have the following observations:
\begin{itemize}
  \item In general, the performance of MAR and MARS are both better than the basic CML in all cases. 
  The performance gains of MAR over CML (\emph{Imp1.}) on the four datasets range from 8.60\% (achieved on BookX with $K$=6) to  54.32\% (achieved on Ciao with $K$=4). The corresponding performance gains of MARS over CML (\emph{Imp2.}) ranges from 32.90\% (achieved on Delicious with $K$=3) to 68.66\% (achieved on Ciao with $K$=4). Such results are consistent with those in Table \ref{tab:exp-all}, showing the effectiveness of both our proposed techniques.
  \item The performance gains of MARS over MAR (\emph{Imp3.}) on four datasets fluctuate, ranging from 1.04\% (achieved on Delicious with $K$=3) to 28.46\% (achieved on BookX with $K$=6), showing the enhancement brought by spherical optimization regarding both performance and robustness. Interestingly, the improvements of MARS over MAR are most significant when the improvements of MAR over CML are small.  Such observations strongly indicate that the spherical optimization is more useful in the more difficult situations, which corroborates our conjecture regarding the weak embedding norm constraints of MAR. 
  \item On all datasets, increasing the number of spaces $K$ leads to larger performance gain of MAR over CML, especially when $K$ is small. However, after the optimal values (often 3 or 4), the improvements start to drop a bit, probably due to insufficient training and overfitting. The improvements of MARS over CML largely follow the trend of MAR, \ie, both frameworks tend to excel with similar $K$'s across four datasets, while MARS does yield more stable improvements. In practice, $K$ should be set according to the complexity of the datasets, and 3 or 4 could be the rule-of-thumb.
\end{itemize}

\begin{figure*}
{
    \subfigure[Item embedding learned by CML]{
    \includegraphics[width = 0.29\linewidth]{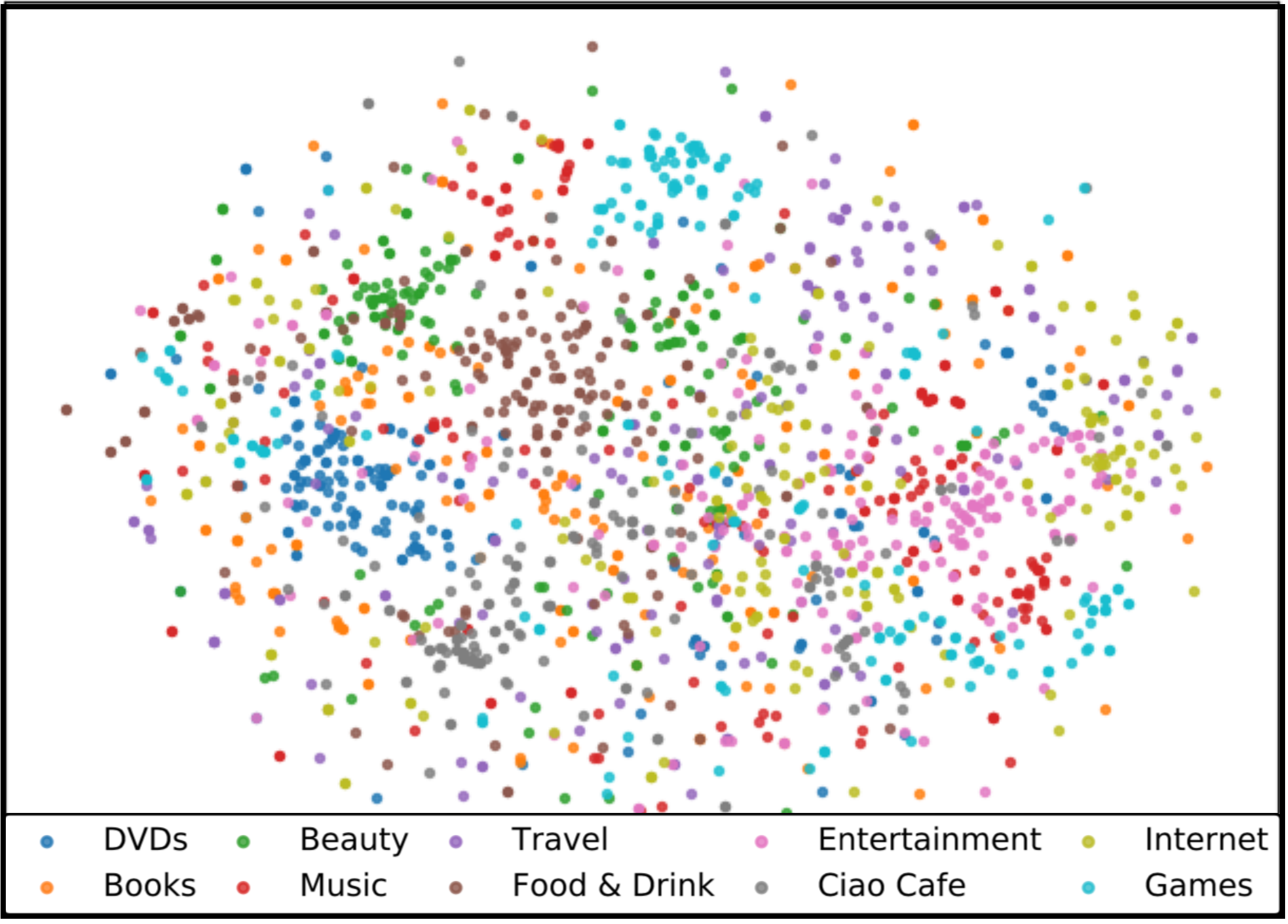}
    }
    \subfigure[Item embedding learned by MAR]{
    \includegraphics[width = 0.29\linewidth]{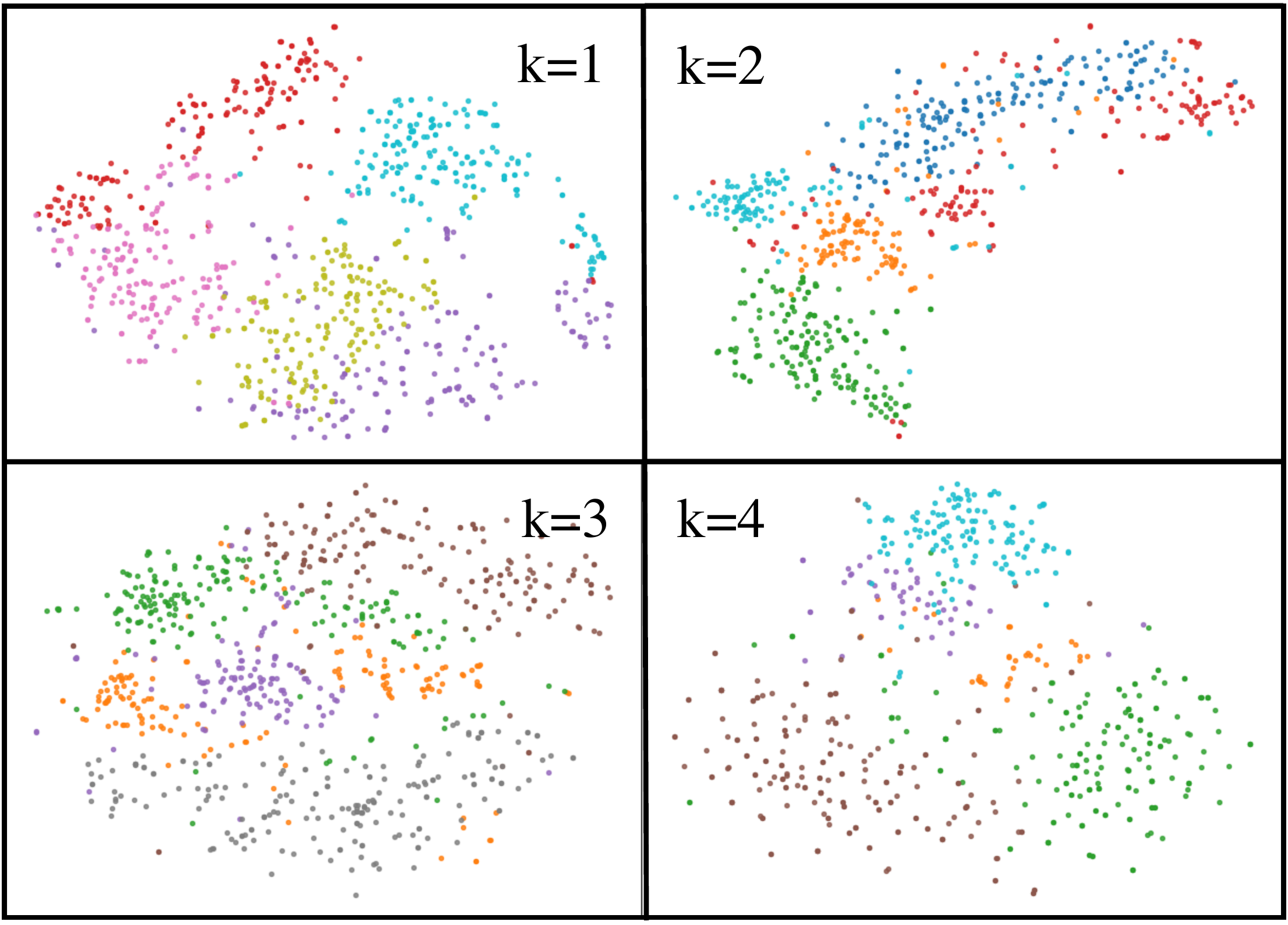}
    }
    \subfigure[Item embedding learned by MARS]{
    \includegraphics[width = 0.29\linewidth]{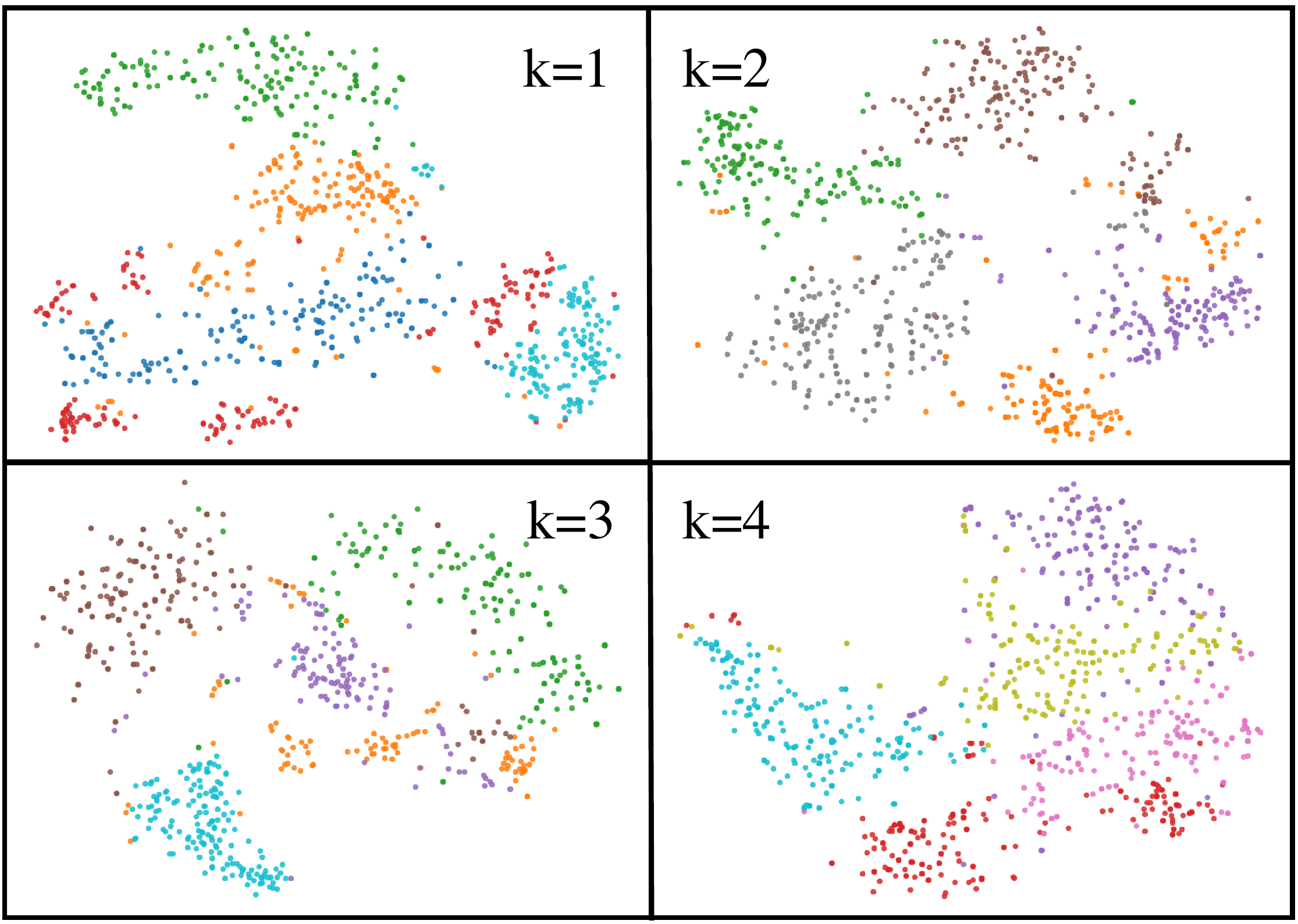}
    }
    \caption{Visualizations of item embeddings learned in a single metric space, multi-facet Euclidean spaces and multi-facet spherical spaces.}
    \label{fig:vs}
}
\end{figure*}

\subsection{Major hyperparameter study (RQ3)}
\label{sec:rq3}
Our proposed frameworks mainly introduce two additional hyperparameters $\lambda_{pull}$ and $\lambda_{facet}$ to control the weight of the ``pulling'' regularizer and facet-separating regularizer, respectively. Here we show how these two hyperparameters impact the performance and clarify how to set them. Due to space limitation, we only show the results with nDCG. There are a few other hyperparameters in the losses inspired by existing works (\ie, $\alpha$ in Eq.~\ref{equ:sim} and $\beta$ in Eq.~\ref{equ:nega}), which are all set to the suggested default values ($\alpha$=0.1, $\beta$=0.8) without tuning.

Firstly, we show the model performance with varying $\lambda_{pull}$. The regularizer $ \mathcal{L}_{pull} $ can pull positive items close towards the corresponding users in the metric spaces. If $\lambda_{pull}$ is too small, the interactions between positive users and items will likely be weakened.
However, too large $\lambda_{pull}$ will likely cause the model to overfit. 
The results are shown in Figure \ref{fig:pull}. We found that the optimal $\lambda_{pull}$ values on Delicious, Lastfm, Ciao and BookX datasets to be about 0.1, 0, 0.001 and 0.1, respectively. Therefore, MARS is reasonably sensitive to $\lambda_{pull}$, as the performance peaks at different values, but is always better than the best baseline. In the range of [0,1], the optimal parameters can be obtained by slight tuning.

Secondly, for hyperparameter $\lambda_{facet}$, the optimal $\lambda_{facet}$ for different datasets is consistently 0.01, as shown in Figure \ref{fig:facet}. In all settings of $\lambda_{facet}$, the performance of MARS again is always better than the best baseline.
In particular, we observe the effectiveness of $\mathcal{L}_{facet}$ as increasing $\lambda_{facet}$ always leads to improved performance when the values are small. However, further increasing it beyond the optimal value always makes the performance worse. In practice, $\lambda_{facet} = 0.01$ seems to be the rule-of-thumb.

\begin{table}[]
    \centering
    \caption{Top-5 categories with proportions in the different embedding spaces of MARS.}
    \vspace{-5pt}
    \begin{tabular}{c  c c   c   c c}
         \hline
         \hline
          &  category & prop (\%) &  & category & prop (\%) \\
         \hline
         \multirow{4}{*}{$k$=1} & DVDs & 10.38 &  \multirow{4}{*}{$k$=2} & Ciao Cafe & 16.07 \\ 
         & Beauty & 8.97 &  & Food \& Drink & 13.85 \\
         
         & Music & 6.65 & & Beauty & 10.34 \\
         
         & BookX & 6.32 & & BookX & 9.34\\
         
         & Games & 5.19 & & Travel & 8.31 \\
         \hline
         & \multicolumn{2}{l}{\textit{House man}} &
         & \multicolumn{2}{l}{\textit{Internet celebrity}} \\
         \hline
         \multirow{4}{*}{$k$=3} & Beauty & 10.45 & \multirow{4}{*}{$k$=4} & Internet & 8.89 \\ 
         & Food \& Drink & 10.45 &  & Entertainment & 8.39 \\
         
         & Games & 5.64 & & Travel & 8.34 \\
         
         & BookX & 5.32 & & Games & 7.34\\
         
         & Travel & 5.22 & & Music & 6.93 \\
         \hline
         & \multicolumn{2}{l}{\textit{College student}} & 
         & \multicolumn{2}{l}{\textit{Software engineer}} \\
         \hline
         \hline
         
    \end{tabular}
    \label{table:k}
    \vspace{-5pt}
\end{table}

\begin{table}[]
    \centering
    \caption{Examples of user profiles modeled by MARS.}
    \vspace{-5pt}
    \begin{tabular}{c|ccl}
        \hline
        \hline
          User & k & $\theta^k_u$ & Interacted categories: interaction number \\ 
         \hline
         \multirow{4}{*}{{Bob}} 
         &k=1 & 0.56 & DVDs: 55; Games: 25; $\dots$  \\
         & k=2 & 0.05 & Ciao Cafe: 61; $\dots$\\
         & k=3 & 0.04 & Food \& Drink:43;  $\dots$\\
         & k=4 & 0.35 & Software: 45; Music: 15 $\dots$\\
        \hline
         \multirow{4}{*}{{Mary}} 
         & k=1 & 0.13 & DVDs: 103; BookX: 60; $\dots$ \\
         
         & k=2 & 0.70 & Ciao Cafe: 99; Beauty: 58; Music: 42;  $\dots$ \\

         & k=3 & 0.05 & Travel: 12; $\dots$ \\

         & k=4 & 0.12 & Internet: 65; $\dots$  \\
         
         \hline
         \hline

    \end{tabular}
    
    \label{table:case}
    \vspace{-5pt}
\end{table}

\subsection{Multi-facet case study (RQ4)}
To demonstrate the advantages of the multi-facet recommendation over single-space systems,
we visualize the embedding vectors learned by CML, the proposed MAR and the proposed MARS on the Ciao dataset as shown in Figure \ref{fig:vs}. CML, MAR and MARS use the same color spectrum to represent different item categories (ground-truth given in Ciao).
In CML, each item corresponds to only one learnable vector in a single space, whereas in MAR and MARS, each item corresponds to vectors in the different facet-specific spaces.
From Figure \ref{fig:vs}, it is hard to find regularity in the distribution of items from different categories in the single embedding space learned by CML. However, the items from different categories are well separated in the multiple embedding spaces of MAR, and the embedding spaces do include different categories of items and distribute them differently. Moreover, the different categories of items are even better organized by MARS, with smaller intra-category distances and larger inter-category distances. The visualizations clearly demonstrate the advantages of MAR and MARS in capturing the multi-facet item property. 

To provide more insights, we further demonstrate the top-5 categories in each facet-specific space (Table \ref{table:k}), and two example user profiles (Table \ref{table:case}), both retrieved based on the user-item interaction data and the user-facet weights learned by MARS.
It is interesting that in Table \ref{table:k} we are able to manually assign some user stereotypes to the learned implicit facets, such as \textit{Internet celebrity} to the space of $k=2$ and \textit{software engineer} to the space of $k=4$, based on their top concerned item categories. 
Moreover, in Table \ref{table:case} we are able to profile two random users (with fake names) as combinations of the stereotypes, and they interact with different items with different roles in the corresponding facet-specific spaces.
Note that, the exact stereotype labels we create here are not perfectly accurate due to the implicit nature of facets, but they nonetheless provide valuable insight into the meaningful representative user types directly extracted from the implicit feedback data in an unsupervised fashion, which provides potential for further user profiling and personalization.
By projecting users and items into different implicit facet-specific spaces, MAR and MARS can learn the fine-grained multi-facet preference of users and property of items, which provides extra knowledge regarding not fully leveraged by existing recommendation systems. 

\section{Conclusion}
In this paper, we propose MARS for recommendation with \carl{multi-space user/item embedding, which can effectively resolve the potential conflicts caused by multi-facet user preferences and item properties}.
Specifically, MARS learns multiple related but dissimilar metric spaces, each of which aims to capture user-item interactions through a particular implicit facet. 
The learning of MARS does not require any external knowledge or algorithm, but only relies on the implicit feedback data. 
We demonstrated the superior performance of MARS in recommendation through extensive experiments, and showcased its effectiveness in multi-facet user/item modeling through insightful case studies.

In the future, it would be interesting to further explore different projections of multi-facet embeddings (\eg, nonlinear), as well as dynamically learn the radiuses of different facet-specific spherical embedding spaces. It is also important to closely study the behavior of MARS regarding the so-called difficult users and items in controlled experiments (such as with users and items grouped based on the number of interactions). Finally, it is potentially useful to infer clusters and attributes of users and items based on the learned MARS model, and utilize them to support other related downstream tasks like user/item segmentation and profiling.

\section{Acknowledgments}
This work was supported in part by the National Key R\&D Program of China (No. 2018YFB1403001) and sponsored by CCF-AFSG Research Fund.

\bibliographystyle{abbrv}
\bibliography{MARS}

\end{document}